\documentclass[pre,showpacs,preprint]{revtex4-1}
\usepackage{amsmath,amssymb}
\usepackage{rotating}
\usepackage{color}
\usepackage[utf8]{inputenc}
\usepackage{graphicx} 
\graphicspath{{./figures}}
\usepackage{epstopdf}
\usepackage{xcolor}
\usepackage{ulem}

\usepackage[colorlinks = true,
            linkcolor = blue,
            urlcolor  = blue,
            citecolor = blue,
            anchorcolor = blue]{hyperref}

\begin{document}

\title{Yukawa Friedel-Tail pair potentials 
for warm dense matter applications}

\normalsize

\author
{
 M.W.C. Dharma-wardana}
\email[Email address:\ ]{chandre.dharma-wardana@nrc-cnrc.gc.ca}
\affiliation{
National Research Council of Canada, Ottawa, Canada, K1A 0R6
}
\author
{Lucas J. Stanek}
\affiliation{Department of Computational Mathematics, Science and Engineering, 
Michigan State University, East Lansing, Michigan 48824, USA}
\author
{Michael S. Murillo}
\affiliation{Department of Computational Mathematics, Science and Engineering, 
Michigan State University, East Lansing, Michigan 48824, USA}

\date{\today}
\begin{abstract}
Accurate equations of state (EOS) and plasma transport properties are essential
 for numerical simulations of warm dense matter encountered in many high-energy-density
 situations. Molecular dynamics (MD) is a simulation method that generates EOS and
 transport data using an externally provided potential to dynamically evolve the
 particles without further reference to the electrons. To minimize computational cost,
 pair potentials needed in MD may be obtained from the neutral-pseudoatom
 (NPA) approach,  a form of single-ion density functional theory (DFT), where
 many-ion effects are included via ion-ion correlation functionals. Standard
  $N$-ion DFT-MD provides pair potentials via the force matching technique but at
 much greater computational cost. Here we propose a simple analytic model
 for pair potentials with physically meaningful  parameters based on a Yukawa form
 with a thermally damped Friedel tail (YFT) applicable to systems containing free
 electrons. The YFT model accurately fits NPA pair potentials or the non-parametric
 force-matched potentials from $N$-ion DFT-MD, showing
 excellent agreement for a  wide range of conditions.
 The YFT form provides accurate extrapolations of the NPA or force-matched potentials for small
 and large particle separations within a physical model. Our method can be adopted
 to treat plasma mixtures, allowing for large-scale simulations of multi-species
 warm dense matter.
\end{abstract}
\pacs{52.25.Jm,52.70.La,71.15.Mb,52.27.Gr}
\maketitle

\section{Introduction}
\label{intro.sec}
Studies in warm dense matter (WDM) systems straddle an intermediate regime of
 matter  extending from cold solids to hot plasmas, at arbitrary compressions
 and states of ionization
\cite{GaffneyHDP18, Grabowski20, Hungary16, cdw-cpp15, NgIJQC12}.
The ions are usually modeled as classical particles whereas the electrons are treated
 quantum mechanically \cite{stanton2018multiscale}. Typical WDM systems often fall
 into regimes of average mass density $\bar{\rho}$ and temperature $T$ where reliable
 experimental data do not exist or are not easily available, as in most astrophysical
 and geophysical applications. The same issue arises in high-energy-density
 applications where energy is deposited in femto-second time-scales producing systems
containing multi-species quasi-equilibria associated  with multiple
 temperatures~\cite{NgIJQC12,CDW-cm-91,ELR98}. The electrons
 and holes in semi-conductor nanostructures also behave like WDM systems due to the
 small effective masses of the charge carriers~\cite{lfc1-dw19}.
The electronic and ionic interactions in these finite-temperature systems are such
 that interparticle effects are strong; hence simple models of the effective
 pair potential fail when applied to such many-particle systems~\cite{Stanek21}. 

Furthermore, the equations of state (EOS)~\cite{GaffneyHDP18} and
transport properties ~\cite{Grabowski20} of WDM systems
 require the tabulation of many data points and the available many-body quantum
 techniques, e.g., density functional theory (DFT), path-integral Monte Carlo (PIMC)
methods etc., become computationally expensive.
Here we aim to leverage results of DFT based $N$-ion simulation
 codes~\cite{VASP, ABINIT} to (i) validate the functional form of physically motivated 
models to represent pair interactions, and (ii) elucidate the regions of
 physical space for which simpler models are valid. Using validated pair
potentials, we avoid carrying out repeated (``on-the-fly") $N$-ion
 DFT calculations as the ion system is evolved using ``classical molecular dynamics''.
 Here, ``classical molecular dynamics" implies that the electronic structure calculation
 needed for ion interactions is pre-computed. We emphasize that this pair potential
 description does {\it not} neglect many-ion effects as explained below. The relationship between
the pair energy and the pair potential, and how higher-order corrections are
included, are considered in Sec.~\ref{disc.sec}. The standard DFT-MD procedure
 that determines the electronic structure of an $N$-atom sample of the material
via Kohn-Sham-Mermin finite-temperature  DFT (e.g., as available in computational
 packages~\cite{VASP,ABINIT}), together with the use of classical MD to evolve
 the ions to equilibrium  will be referred to here as ``QMD" for brevity.

The computational cost associated with $N$-ion DFT can be reduced by using
a single-ion DFT approach~\cite{DWP82,eos95,Hungary16,cdw-pop21} where
an ion-ion exchange-correlation (XC) functional is used to reduce the
many-ion problem to that of a single ion such that the material is a simple
 superposition of {\it pseudoatoms}. This does  not mean that a
simple superposition of actual atoms is used, as may be possible 
in weakly coupled systems.
This ``single-ion'' DFT approach to the many-ion, many-electron
problem is referred to here as the neutral-pseudoatom (NPA) model. 
The NPA calculations, being similar to average-atom (AA) calculations,
are computationally rapid in comparison to QMD calculations,
 and usually produce EOS data and pair potentials that are in good agreement with
QMD results~\cite{Stanek21,cdwSi20,Hungary16}. The NPA calculations also directly yield single-ion
properties like the mean-value of the charge ($\bar{Z}$) carried by ions,
 electron-ion scattering cross sections, or electron-ion pseudopotentials that
 are useful for transport calculations~\cite{Grabowski20}. Note that
 there are many 
AA models~\cite{SternZbar07,Rosznyai08,FausAA10,Murillo13,StaSau13,StaSau16}
that are similar in spirit but differ in their details. Examples include
 different choices of functionals used to treat exchange-correlation and kinetic energy,
and more significantly, the boundary conditions used in defining the AA
which may be confined to a Wigner-Seitz sphere or to a much larger volume~\cite{DWP82}.
The particular NPA model used here is described in detail
 in Refs.~\cite{cdwSi20,cdw-Carbon10E6-21,eos95} 
and will be referred to here generically as ``NPA." 

Recently, machine learning (ML) methods have been employed to accelerate the
 calculation of multi-center interatomic potentials (MCPs) whose simplest member
 is the pair potential.  One of the simplest ML methods is force
 matching \cite{ercolessi1994interatomic} which aims to minimize a loss function
 constructed from interionic forces from a QMD dataset and
 a predetermined force law 
\cite{Stanek21, Brommer2007Potfit,Brommer2006Potfit, Brommer2015Potfit}.
 Such a loss function may take the form
\begin{align}
    L(\xi) =  \sum_{k=1}^{3M}w_k^F(F_k(\xi) - F_k^0)^2 &+ 
 \sum_{\ell=1}^M w_\ell^E (E_\ell(\xi)
    - E_\ell^0)^2\nonumber\\ &+ \sum_{m=1}^{6M} w_m^S(S_m(\xi) - 
S_m^0)^2,\label{eq:LF}
\end{align}
where $M$ denotes the number of particle coordinate configurations, $F_k(\xi),
 E_\ell(\xi)$, and $S_m(\xi)$ are the force component, energy, and stress tensor
 component from the target interaction potential parametrized by $\xi$ with
 weights $w_k^F, w_\ell^E,$ and $w_m^S$. The reference force component, energy,
 and stress component, $F^0_k, E^0_\ell$, and
 $S^0_m$, are
 determined from high-fidelity calculations such as Kohn-Sham DFT MD. The last
 two terms in Eq.~(\ref{eq:LF}) can be thought of as \textit{regularizers} of the
 loss function that restrict the functional from of the optimized interaction
 potential. In the following, all of our force-matched (FM) potentials assume a
 loss function given by Eq.~(\ref{eq:LF}) with $w_\ell^E = w_m^S = 0$. More advanced
 machine learning approaches allow for greater than pair or three-body
 interactions by including density-density correlations \cite{thompson2015spectral}. 
Deep neural networks have also been employed to learn the QMD potential energy
 surface where translational, rotational, and permutational symmetries are 
enforced \cite{zhang2018deep,DeepLearning-QZeng21}. 
In all of the aforementioned ML methods the MCP is obtained for use in classical
MD simulations. In the MCP approach the electron sub-system
 is suppressed, and hence higher-order (e.g, 3-body,
4-body) potentials are needed to capture the electronic
energies that are not explicitly included; the philosophy is similar to that of the
 classic Stillinger-Weber potentials \cite{SW85}. 

In contrast, the NPA approach uses only pair potentials, but explicitly includes
 electron contributions to the total free energy  directly available from the
NPA calculation, as indicated in more detail in Sec.~\ref{disc.sec}.
The NPA uses the electron density to calculate energy
contributions from the electron subsystem whereas the
MCP method tries to embody them in the MCPs that carry ``electronic bonding
 information."  This is claimed to allow the MCP method to deal with structural
 defects (e.g., dislocations in metals) ``on the fly," without resorting to
 explicitly handling electronic effects  arising from such inhomogeneities. There
is however no underlying principle in physics which asserts that there
 exists a general classical MCP that determines the physics of general
 electron-ions systems in the way claimed by ML; the only available principle, due
to Hohenberg and Kohn, is that there exists a one-body density functional that is
 universal.

An advantage of ML methods is that the functional form of the ion-ion pair
 interaction potential is allowed to be determined by data, eliminating any bias
 associated with assuming too simple a functional form based on a limited physical
model. On the other hand, the lack
 of a functional form prevents accurate recovery of asymptotic forms imposed
 by the physics of these systems.
The data-driven nature of these ML  methods is also one of their main
 limitations: the pair interaction potential will only be as accurate as the data
 used in the training process. It is possible that the QMD dataset may consist of
 too few time steps, resulting in a poor representation of the potential energy
 surface. The QMD dataset may also contain errors that originate
 from simulations of a small number of particles (referred to herein as
 {\it finite-size effects}). Quantifying errors from finite-size effects is 
 critically important near phase transitions where the cooperative effect of a
 large number of particles occurs. 

Furthermore,
as very complex $N$-atom structures become possible with large-$N$ simulations,
increasingly  more advanced XC functionals  become imperative. It has been shown
that  numerical results using different XC functionals at different levels of the
``Jacob's ladder'' may give significantly different results in
 such calculations~\cite{Remsing17}.

To mitigate the  concerns associated with dataset
 inadequacies, we present a parametrized analytic form of a pair potential
 {\it based on a general physical model} that is found to
 be able to fit  either to potentials extracted  from QMD  simulations or given
 by NPA calculations. If the QMD dataset is expected to be comprehensive enough to
 describe the potential energy surface, the analytic fit form will correct
 for finite-size effects at both small and large interatomic distances.

We note that the NPA is a DFT method where both the electron many-body problem
 {\it and} the ion many-body problem are jointly reduced to equivalent
non-interacting one-body problems. This is done by relegating higher-order effects
 beyond mean-fields into exchange-correlation (XC) functionals; these depend only
 on the electron one-body density and the ion one-body density. Since a central ion
 is used as the origin of the NPA calculation, the one-body ion-density becomes
 effectively a two-body ion density and correlations beyond the two-body potential
 appear in the ion-ion XC-correlation functionals; classical ions have no exchange,
and the correlations used in the NPA for solids, liquids, or hot plasmas
 are those brought in by the  Ornstein-Zernike procedure. Thus, while the
ion-ion XC-functional is strongly non-local, the electron-electron XC-functional
used in the NPA is the finite-$T$ functional given by Perrot and
 Dharma-wardana~\cite{PDWXC}, within the local-density approximation. 
Due to the simplicity of the electron density profile in the NPA (which is a
 single-center screened ion), the local-density approximation is found to be sufficient.  

In the following we treat simple fluids like liquid Al
as well as complex liquids like liquid C, not only at high temperatures, but 
also at low temperatures where the static
structure factors $S(k)$ and pair-distribution functions (PDFs) $g(r)$ may be thought
 to depend on bond-angle
 distributions and other complex correlation effects. However, density functional theory is a
 theory of the ground
 state energy for systems at $T=0$ and a theory of the free energy
for finite-$T$ systems. These depend only on the time-averaged picture of
 the fluid which is uniform as we are not treating nanostructures or solids.
 Hence there is no need to include the detailed instantaneous bond-orientational
 details that are generated in standard $N$-ion
 simulations, only to be averaged out using millions of such calculations. 
The static (i.e., long-time average) correlations beyond
 the two-body terms are brought into the NPA calculations
 via the Ornstein-Zernike procedure which requires the
use of an ion-ion XC functional, as shown in Ref.~\cite{DWP82}.
The energy  effect of such correlations is taken care of in the NPA through the 
 ion-ion XC functionals, as discussed in Ref.~\cite{DWP82}.
 Furthermore, the energy and free energy
 are functionals of just the {\it static} one-body densities. Unlike in $N$-ion DFT
 simulations, calculations using the NPA method will not yield instantaneous bonding details
 and ``snap shots'' of the fluid, although the {\it static} PDFs and static
 structure factors become readily available. However, as shown by Hafner~\cite{Hafner89}
 for liquid As, a simple pair potential can be used to provide  considerable information
 on bond-angle distributions by using the pair potential in a classical simulation.

In this study, we focus on the pair potentials obtained from the NPA model, and on
force-matched potentials extracted from QMD where they are available.  They
are of interest for simulating physical systems such as liquid metals, WDM systems, and strongly
 coupled plasmas where the primary input to classical MD
 simulations is the pair potential~\cite{cdw-aers83,chenlai92,Utah12,VorbGeri-Pots13,
whitley15,Stanek21}. Specifically, we use the NPA model to generate pseudopotentials
 and pair potentials needed for computations of physical properties of
 finite-temperature  electron-ion systems of uniform density, and show that a simple
 physically meaningful parametrization of the potentials exists for them. The simplest
 method, based on a linear-response approach to the pseudopotentials,
is found to be applicable for a wide range of densities and temperatures.
Examples where linear-response fails are common for expanded metals,
liquid (which we denote as $l$-) C, and $l$-transition metals at low $\bar{\rho}$ and
 $T$~\cite{Stanek21}. In such cases special procedures and sophisticated XC-functionals
 are needed, even in standard $N$-atom DFT. 

In this study, we provide pair potentials for simple
fluids like $l$-Al and complex fluids like $l$-C whose QMD simulations
show the existence of transient covalent bonds. Nevertheless, it was
shown in detail that a simple-metal model is applicable to the
thermodynamic and structural properties of $l$-C and
$l$-Si, and even to their phase transitions~\cite{cdw-carb22,cdwSi20}.  
These results may seem to go against ``chemical intuition'' but are consistent with
 well established physics-based insights on the ionic correlations
 in simple and complex liquid metals and solids~\cite{Hafner87}.

The remainder of this manuscript is organized as follows. First, in
 Sec.~\ref{sec:NPA_PP}, we present a brief overview of the NPA model and pair
 potentials. We discuss the theory of NPA pseudopotentials which originates from the
 Kohn-Sham DFT electron density. We also describe how pair potentials stem
 from linear-response  theory and their implementation in the NPA model.
 The NPA pair potentials are shown to agree closely with force-matched QMD
 pair potentials which are available for length scales limited to the size
 of the simulation box used in QMD. Then, in Sec.~\ref{parampot.sec},
  we introduce a simple but generally applicable  parametrization scheme
 for pair  potentials in homogeneous metallic systems.
 The proposed parametrization is consistent with a simple physical picture
  based on screened Coulomb interactions of the Yukawa type, and includes thermally
 damped Friedel  oscillations as they become long-ranged and important
 for $T<E_F$. In Secs.~\ref{car-pot.sec} to \ref{li-pot.sec} we present fits to
 our parametrization scheme using pair-potentials from NPA calculations for
  $l$-C, $l$-Al, and $l$-Li.  We show that our parametrized scheme
 accurately reproduces structural quantities obtained from QMD or PIMC calculations
 and can be used to  greatly reduce errors incurred
 from finite-size effects in QMD simulations. 

\section{\label{sec:NPA_PP}The NPA pseudopotentials and pair potentials}
The NPA calculation self-consistently
 generates the Kohn-Sham one-body electron density distribution $n(r)$ around
a nucleus placed in the appropriate environment in the fluid where an
all-electron calculation is used. For simplicity of exposition here we
 consider the case where the bound electrons
are well localized inside the Wigner-Seitz sphere of radius $r_{\rm ws}$
 associated with the ion. Then, the electron distribution can
be unambiguously written as a sum of a bound electron part denoted as $n_b(r)$
and a free electron part denoted as $n_f(r)$, while the issues arising from the more
general case are discussed in Appendix~\ref{sec:bound_free}. The free electrons are {\it not}
confined to the Wigner-Seitz sphere (as in many AA models), and occupy the whole of 
space represented by a large ``correlation sphere'' of radius $R_c$ which is
some five to ten times $r_{\rm ws}$, chosen to ensure that all interparticle
 correlations are sufficiently small by $r=R_c$, i.e,
all pair distribution functions (PDFs) $g(r)\to 1$ as $r\to R_c$. The electron
chemical potential is the non-interacting value used in DFT and
need not be re-assigned as in many AA models that use a Wigner-Seitz sphere to define
and AA. The free-electron density $n_f(r)$ is used to construct a
 pseudopotential which is then employed to construct a pair potential. 
If a metallic model with strong screening is assumed, the above steps
can be carried out within linear-response theory, giving
rise to simple local (i.e., $s$-wave) pseudopotentials  that contrast with the
non-local non-linear pseudopotentials supplied with QMD
codes \cite{VASP,ABINIT}.

At the end of an NPA calculation we have the free-electron density distribution
 $n_f(r)$ around the ion of mean charge $\bar{Z}$,
where the average (uniform) free-electron density $\bar{n} = \bar{Z}\bar{\rho}$ 
as required by charge neutrality. The evaluation of $\bar{Z}$ used in the
NPA is both an atomic physics calculation and a thermodynamic ionization-balance
calculation that satisfies the Friedel sum rule and  charge neutrality.

\subsection{The pseudopotential}
Finite-$T$ pseudopotentials based on NPA-type models were first used by
Dharma-wardana and Perrot~\cite{DWP-carb90} and Perrot~\cite{Perrot90}, following
the zero-$T$ analogues used in the physics of metals. The pseudopotential is the
 potential exerted by the unscreened ion
of charge $\bar{Z}$ made up of the nucleus plus its core-electrons. The direct output of
the NPA calculation is the full non-linear response of the electron fluid to the central
 nucleus together with the ion distribution whose $g(r)$ is modeled by  a simple spherical
 cavity of radius $r_{\rm ws}$. The latter has been self-consistently
adjusted to satisfy the NPA equations, as described in Ref.~\cite{cdwSi20}.
Then, the free-electron pileup $\Delta n_f(r)$ caused purely by the nucleus and its
bound core is obtained by correcting for the effect of the $g(r)$ modeled as a cavity 
according to the method discussed in Refs.~\cite{cdwSi20} and~\cite{eos95}.
That is, we use the cavity-corrected free-electron charge pileup
$\Delta n_f(r)=n_f(r)-\bar{n}$ and its Fourier transform $\Delta n_f(k)$
to construct the electron-ion pseudopotential (in Hartree atomic units),
\begin{eqnarray}
\label{uei.eqn}
U_{ei}(k)&= &\Delta n_f(k)/\chi(k,r_{\rm{s}},T)\nonumber\\
\label{formfac.eqn}
         &\equiv &-\bar{Z}V_kM_k.
\end{eqnarray}
The negative sign in the second equation arises from the electron chagre $e=-1$.
 Here $V_k=4\pi/k^2$ and $\chi(k,r_{\rm{s}},T)$, which we abbreviate as $\chi(k)$, is the
 fully interacting finite-temperature response function of the uniform electron
 fluid at the density $\bar{n}$ associated with the {\it{electron}} Wigner-Seitz
radius $r_{\rm{s}}$. The behavior of the screening function heavily depends on the
 degeneracy parameter $\theta=T/E_F$ where $E_F$ is the Fermi energy.
The details of the calculation of $\chi(k)$ as well as the limitations of this
 pseudopotential are discussed in Refs.~\cite{Stanek21, cdw-pop21}.
Here we note that an all-electron non-linear DFT calculation has been
applied to the electron-nuclear interaction in obtaining $\Delta n_f(r)$,
and linear response is used only in defining a pseudopotential. Thus the
 pseudopotential incorporates the non-linear aspects of the Kohn-Sham calculation.

The domain of validity of the method of constructing pseudopotentials
can be extended considerably if the response function $\chi(k)$ based on
 planewaves is replaced by a form  constructed from the free-electron
Kohn-Sham  eigenfunctions of the NPA calculation itself.
However, in this study we use the simplest form which has been found
to be accurate even for complex fluids like $l$-C and $l$-Si unless
low-density low-temperature systems are considered.

\begin{figure}[t]
\includegraphics[width=.98\columnwidth]{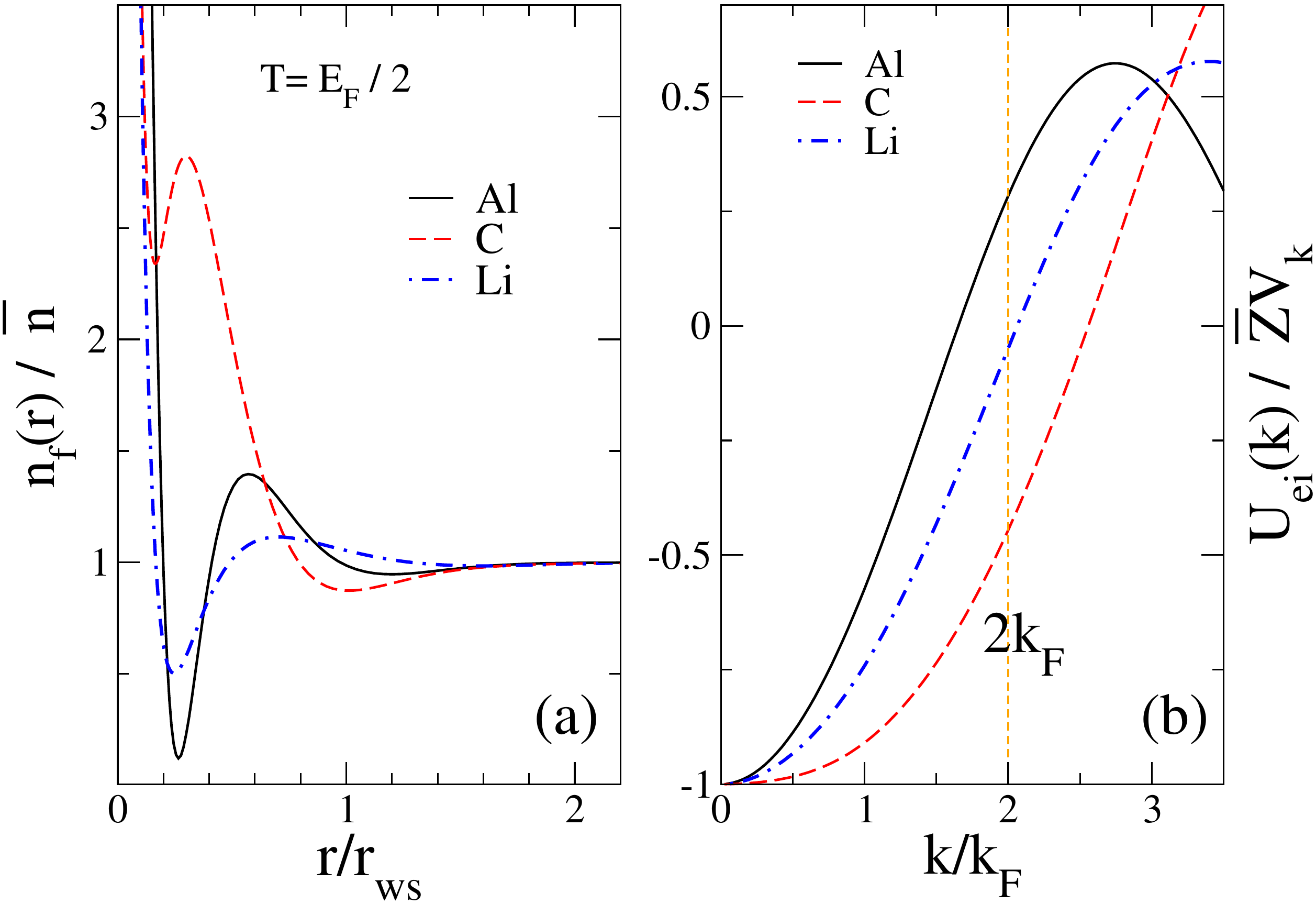}
\caption
{(Color online) Free-electron densities and  pseudopotentials for the elements
 Al, C, and Li. (a) The NPA free-electron density $n_f(r)$ normalized by the
average free-electron density $\bar{n}$ for Al$^{3+}$, C$^{4+}$, and Li$^+$ at
$\theta=T/E_F$=0.5, at  densities $\bar{\rho} = $ 2.7 g/cm$^3$, 3.7 g/cm$^3$, and
 0.513 g/cm$^3$ respectively. Different elements are best compared using scaled
 lengths $x=r/r_{\rm ws}$ and scaled momenta $q=k/k_F$ where $k_F$ is the Fermi
 momentum. (b) The electron-ion pseudopotential $U_{ei}(k)$ via Eq.~(\ref{uei.eqn})
 is given in terms of the bare point-ion potential $\bar{Z}V_k$. That is, the
 form factor  $M_k=U_{ei}(k)/\bar{Z}V_k$ is unity for an unscreened point-ion. The
behaviour of the pseudopotential for values of $k/k_F$ significantly greater
than 2 is unimportant.   
\label{denUei.fig}
}
\end{figure}

\begin{figure}[t]
\includegraphics[width=.98\columnwidth]{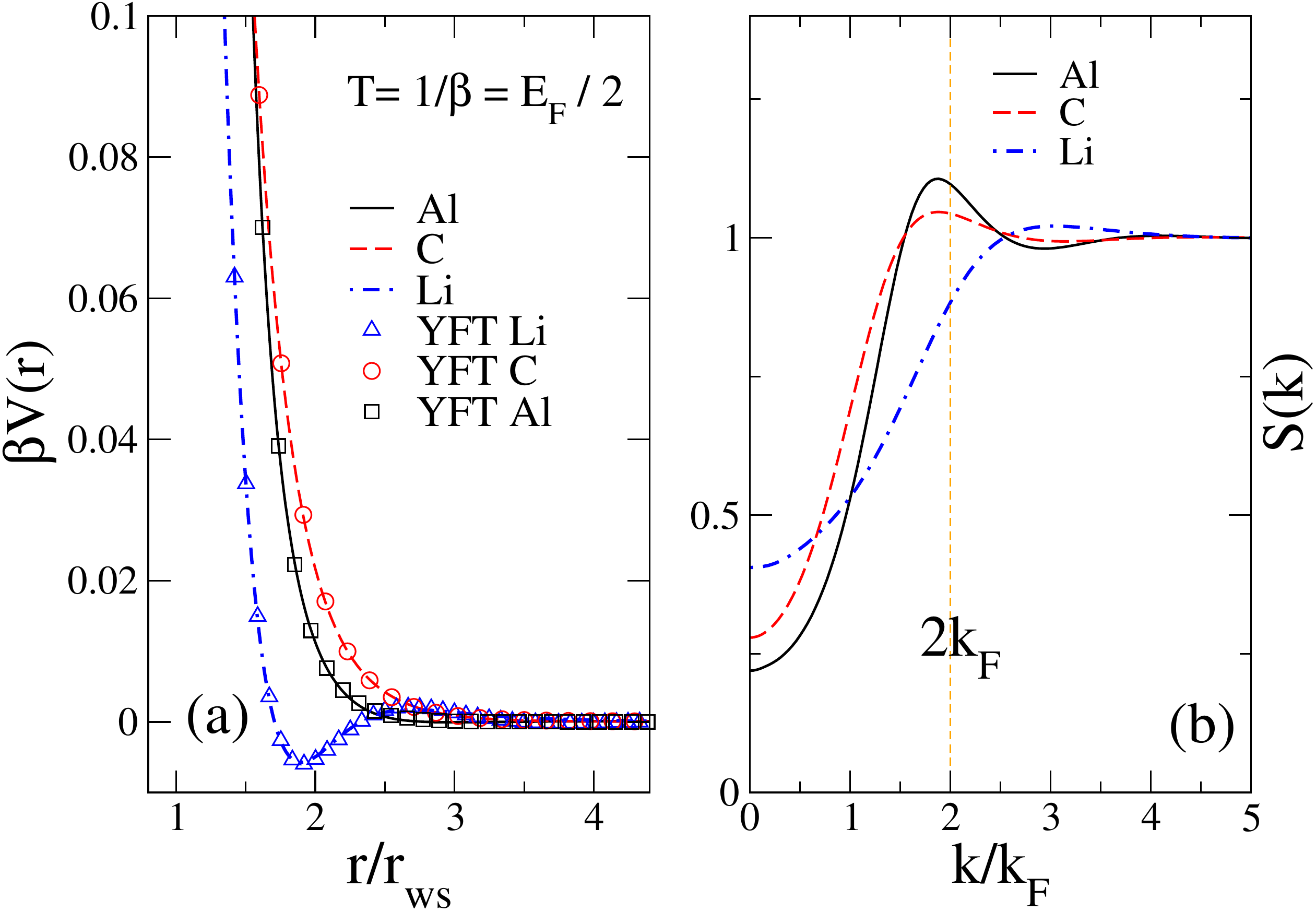}
\caption
{(Color online) Pair potentials $V(r)$ and structure factors $S(k)$ for Al, C, and Li
 at the same conditions given in the caption of Fig.~\ref{denUei.fig}.
(a) The ion-ion pair potentials $V(r)/T$ for C, Al, and Li are computed from the
 pseudopotentials $U_{ei}(k)$ (see Fig.~\ref{denUei.fig}(b)) using Eq.~\eqref{uei.eqn}.
 The $V(r)$ is accurately parametrized
by the YFT form (see Sec.~\ref{parampot.sec}). The YFT fits 
are shown as symbols (triangles, circles, or squares) and data from the NPA calculations
 are shown as solid lines.
(b) Ion-ion structure factors $S(k)$ are obtained from the pair potentials
$V(r)$ using the MHNC equation. At low temperatures when, $\theta <0.25$, the Friedel
 oscillations are significant; the Friedel oscillations 
 become more damped when $\theta = 0.5$.
\label{VrSk.fig}
}
\end{figure}

Examples of the free-electron density $n_f(r)$ and the corresponding
pseudopotential $U_{ei}(k)$ from NPA calculations are shown in
 Fig.~\ref{denUei.fig} for Li, Al, and C; these being examples of ions which
cover the range $\bar{Z}$ = 1 to 4,
for temperatures of interest in WDM studies. To compare the potentials appropriately,
 we have used $x=r/r_{\rm ws}$ as the length scale, and momenta scaled by
the Fermi wavevector $k_F$. The appropriate energy scale for the
electron fluid is the Fermi energy $E_F$, and hence the results
are given for the Fermi temperature $\theta=T/E_F=0.5$ where the degeneracy effects
 are similar in all three materials. Figures~\ref{denUei.fig} and \ref{VrSk.fig} 
demonstrate the basic steps where an NPA calculation proceeds from a
first-principles atomic-physics type calculation of the charge densities
to a determination of the ionic and electronic structure and physical
properties of the complete fluid given the nuclear charge, temperature and
density of the material.

In Eq.~\eqref{formfac.eqn}, the pseudopotential $U_{ei}(k)$ is re-expressed in terms
of the point-ion potential $\bar{Z}V_k$ and the form factor $M_k$
which is unity only for an unscreened point ion. Here $U_{ei}(k)$ is a simple
local $s$-wave potential and we have found it adequate for quantitatively
reproducing results obtained from QMD where advanced non-local
pseudopotentials have been used. The resulting
 pseudopotentials are shown in Fig.~\ref{denUei.fig}(b).
The pseudopotentials can be fitted to a Heine-Abarankov form~\cite{Utah12},
or more detailed parametrizations as provided in
Ref.~\cite{ELR98}.

For completeness we note that these pseudopotentials reduce to popular
variants of the Yukawa potential
~\cite{SunDai-Yuk17,Stanton15,Garavelli91,CDW-RT-81,Rogers70,Yukawa35}
when the form-factor is set to unity and
the $k\to 0$ limit of the Lindhard function is used as
the response function. Then the screened pseudopotential becomes
\begin{eqnarray}
U_{ei}(k)/\epsilon(k) &=& -\bar{Z}/(k^2+k_s^2),\\
\epsilon(k) &=& 1+k_s^2/k^2,
\end{eqnarray}
where the $k\to 0$  limit of the random-phase
approximation (RPA) to the dielectric function $\epsilon(k)$ has to be used.
Such models become accurate in the high-density,
$\theta \to 0$, Gell-Mann - Brueckner limit where the Yukawa potential
becomes the Thomas-Fermi screened potential, and in the low density,
high-temperature limit ($\theta \gg 1$) of Debye-H\"{u}ckel-like systems.
The intermediate case and related approximations are discussed in
 Ref.~\cite{CDW-RT-81}. The form proposed by Stanton and Murillo~\cite{Stanton15}
 is a partial linearization based on the Thomas-Fermi-Weizs\"acker
finite-$T$ density at an atomic nucleus, but still in the long-wavelength ($k\to 0$)
 limit. Finite-$T$ screening at finite-$k$ based on the
homogeneous-electron-gas RPA model is the next step beyond the
Yukawa form and has been studied by Perrot~\cite{perrotRPA91}. 
We have improved on these in the NPA potentials in two ways: (i) we
go beyond the RPA by using a local-field correction that satisfies the
finite-$T$ compressibility sum rule in regard to the response 
function, and (ii) we go beyond the point-ion or Thomas-Fermi-Weizs\"acker models
by employing a form factor and incorporating the {\it full non-linear
Kohn-Sham density response} in the electron-ion interaction $U_{ei}(k)$.
This is done by the simple procedure given in Eq.~(\ref{formfac.eqn}).

\subsection{The pair potential}
\label{pairpot.sec}
Neutral pseudoatoms are non-interacting objects and their free energies are additive
quantities. It is the \textit{unscreened} pseudoatoms (ions of charge $\bar{Z}$ free
 of their screening sheaths) that interact. The \textit{pair potential} is the
 interaction energy between an unscreened pseudo-ion in the presence of another
 unscreened pseudo-ion, at a distance $r$ apart in the medium containing only a
 uniform electron subsystem, while a static (non-responding) neutralizing
 positive-ion background is implicit. This interaction between two unscreened
  pseudo-ions is just their Coulomb interaction moderated by the screening effect
 of the electron subsystem. The pseudopotential was constructed by ``unscreening''
 the pseudoatom density using linear response theory, as in Eq.~(\ref{uei.eqn}).
 Hence the use of linear-response theory to determine the effect of screening on
 the pair interaction is appropriate, and corresponds to second-order perturbation
theory using the pseudopotential $U_{ei}(k)$. 

 Using Eq.~(\ref{uei.eqn}), the ion-ion pair potential in Hartree atomic units is
\begin{eqnarray}
\label{pair.eqn}
V_{ii}(k)&=&\bar{Z}^2V_k+|U_{ei}(k)|^2\chi(k),\\
\label{dft-pair.eqn}
         &=&Z^2V_k+|\Delta n_f(k)|^2/\chi(k,T)
\end{eqnarray}
Equation~(\ref{pair.eqn}) has been used in the theory of liquid metals
since the 1960s but with empirical pseudopotentials. Here we use
the NPA based first principles pseudopotentials to construct the interaction
 between two unscreened pseudo-ions of the same type~\cite{Perrot90}.
The form given in Eq.~\ref{dft-pair.eqn} emphasizes the non-linear DFT
density pileup $\Delta n_f(k)$ and indicates that the linear-response
assumption is contained in using $\chi(k,T)$ as the electron response.

 The interaction between two unscreened
 pseudo-ions of different types immersed in the same electron subsystem, 
at the temperature $T$ is given by,
\begin{equation}
\label{mixedpair.eqn}
    V_{ij}(k)=\bar{Z}_i\bar{Z}_jV_k+U_{ei}(k)U_{ej}(k)\chi(k).
\end{equation}
Here the ions have charges $\bar{Z}_i$ and $\bar{Z}_j$ and pseudopotentials
 $U_{ei}(k)$ and $U_{ej}(k)$. As before, the
 pseudopotential used here is also the simplest local ($s$-wave) pseudopotential; 
 no angular-momentum projected (non-local) forms have been found necessary
 in our studies of uniform WDM fluids or  even for calculations on cubic
 solids using NPA~\cite{HarbEOSPhn17}. In contrast, $N$-atom QMD calculations of
 such systems often employ non-local potentials.  The simple
 pseudopotential in
NPA arises from the fact that the NPA does not attempt to account
 in detail for directional bonding when spherical averages suffice.

The $r$-space form of the pseudopotential is obtained by Fourier transformation,
and hence the calculation of pseudopotentials and pair potentials from the NPA
within linear response is computationally rapid, and provides accurate DFT results for
single-component systems. A discussion on using this NPA approach
 for multi-component systems is presented in Sec.~\ref{mixture.sec} and follows
Ref.~\cite{eos95}. In describing the general method we use a discussion based
 on a single-component system
 with the ions and electrons all in equilibrium at a temperature $T$.
Thus the ion-ion linear response pair potential $V_{ii}(r)$ will be denoted by
$V(r)$ where no ambiguity arises. The pair potentials for systems where
the electron temperature $T_e$ is different from the ion temperature
$T_i$ may be generated as presented in Harbour {\it{et al.}}~\cite{DSF18}.

The pair potentials for Al, C, and Li are displayed in Fig.~\ref{VrSk.fig}(a);
 they are computed from
 Eq.~(\ref{pair.eqn}) using the pseudopotentials shown in Fig.~\ref{denUei.fig}(b).
 The corresponding ion-ion structure factors $S(k)$ are computed from the modified
 hyper-netted-chain (MHNC) equation and are displayed in Fig.~\ref{VrSk.fig}(b).
 We stress that the NPA model holds for arbitrary electron-ion systems ranging
 from low to high temperatures as long as the main assumption of
the validity of linear response is upheld.
 Starting from the pseudopotential, pair potential,
 and structure factor generated from NPA, it is possible to determine the thermodynamic properties
 as well as linear transport properties~\cite{eos95,cond3-17} of the system. For example,
 the EOS and thermodynamics
can be obtained directly from the total free energy. Electrical transport coefficients like
 the electrical conductivity can be computed from Ziman-type formulae from Kubo-Greenwod theory,
 or from Fermi golden-rule type considerations~\cite{Utah12}. Ionic transport properties like
 diffusion and viscosity can be obtained from the Green-Kubo relations by using the the pair
 potentials as an input to classical MD simulations~\cite{Stanek21}.
 
\begin{figure}[t]
\includegraphics[width=.95\columnwidth]{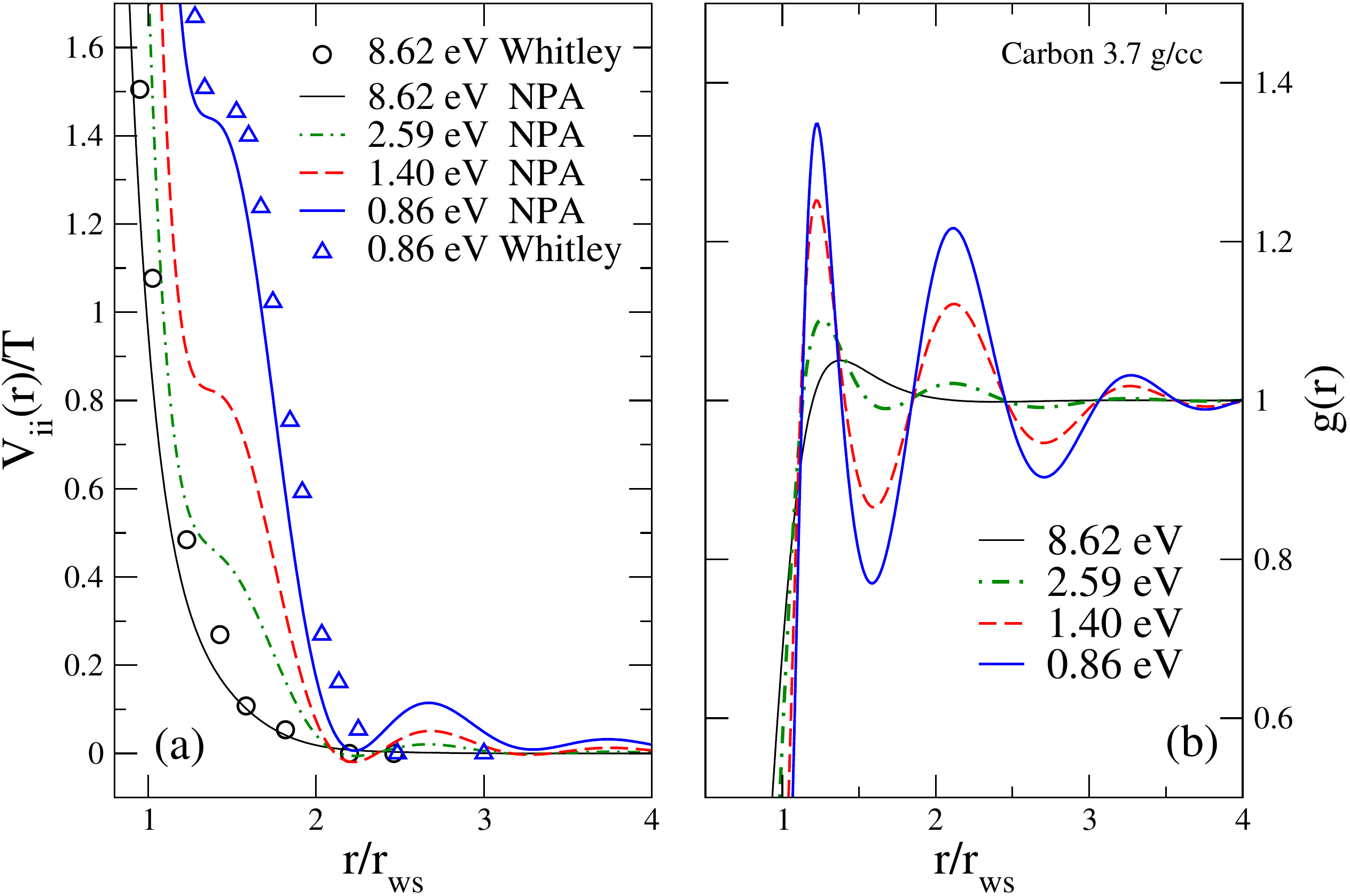}
\caption{(Color online) (a) The QMD pair potentials of $l$-C
 at $\bar{\rho}$ = 3.7 g/cm$^3$ by Whitley {\textit{et al.}}.~\cite{whitley15}
are compared with the NPA linear-response pair potentials,~Eq.~(\ref{pair.eqn}),
 from $T = $ 0.86 eV to 8.62 eV. The QMD  potentials fail to recover the Friedel
oscillations (for $r/r_{\rm ws}>3$) seen in the NPA potentials and may not,
for instance, capture long-range correlations important  at phase transitions.
(b) Two illustrative examples of the YFT fits to the pair-potentials (log-$y$ scale).
 The potentials are nearly Yukawa like at higher $T$, but develop
 Friedel oscillations at lower $T$.\label{Vr.fig}
}
\end{figure}

A comparison of the pair potential obtained with the NPA model and QMD
 calculations for C is displayed in Fig.~\ref{Vr.fig}, showing excellent agreement.
 Many other comparisons were given in Ref.~\cite{Stanek21} also showing reasonable
 agreement for many elements at various average densities and temperatures.
The electrons have been eliminated from such potentials which require
three-center (and higher) terms in the total free energy $F$. In contrast, the NPA 
retains contributions to $F$ from the two-component system of free electrons
and ions, pair interactions, and their XC-contributions to the
free energy that merely depend on the electron-sphere radius $r_s$ and
 $T$ (see Sec.~\ref{disc.sec}).

It is important that the pair potential is accurate in the regime of
radial distances  $r>r_{\rm ws}$ since neighbouring atoms are usually
located in shells that follow the stationary points (minima or inflection
points) of the Friedel oscillations of  the pair potential where possible.
 So, the Kohn-Sham equations of the NPA are solved for
 $R_c\sim 10r_{\rm ws}$ (or 5$r_{\rm ws}$ at higher $T$),
unlike in many AA models where the free-electrons are
confined to the Wigner-Seitz sphere. As the particle correlations become weak
 at  higher temperatures, a smaller $R_c$ can be used in NPA
calculations since Friedel oscillations  become increasingly weak
and die off within a short range.

\section{Yukawa Friedel-Tail potential: A general parametrized form of the
 pair potential}
\label{parampot.sec}
The ion-ion pair potential, given in Eq.~(\ref{pair.eqn}), when Fourier
transformed gives a numerical form for $V(r)$. Since various model forms
based on the Yukawa potential and its  modifications have been studied 
in the context of WDM, it is of interest to examine
the structure of $V(r)$  and provide a general physical model that can
successfully parametrize these potentials accurately, using only a minimum
of parameters that also have a physically clear meaning. 

Such a model consists of a Yukawa-screened ion carrying a long-ranged Friedel tail
referred to in the following as the ``YFT" model. Here, we provide parametrizations
 for Al, C, and Li as examples of YFT potentials. We find that the model
 is useful even in regimes where the NPA technique becomes inaccurate, for example, for low-density
low-temperature $l$-C, as shown by Stanek \textit{et al.}~\cite{Stanek21}. 
The FM low-density low-temperature potentials obtained
 from QMD simulations using Eq.~(\ref{eq:LF}) do not recover many Friedel
 oscillations due to  finite-size effects. However, the FM  potentials can be
 reliably extended by fitting to the  YFT model explicitly given in
 Eq.~(\ref{potform.eqn}). 

\begin{figure}[t]
\includegraphics[width=.95\columnwidth]{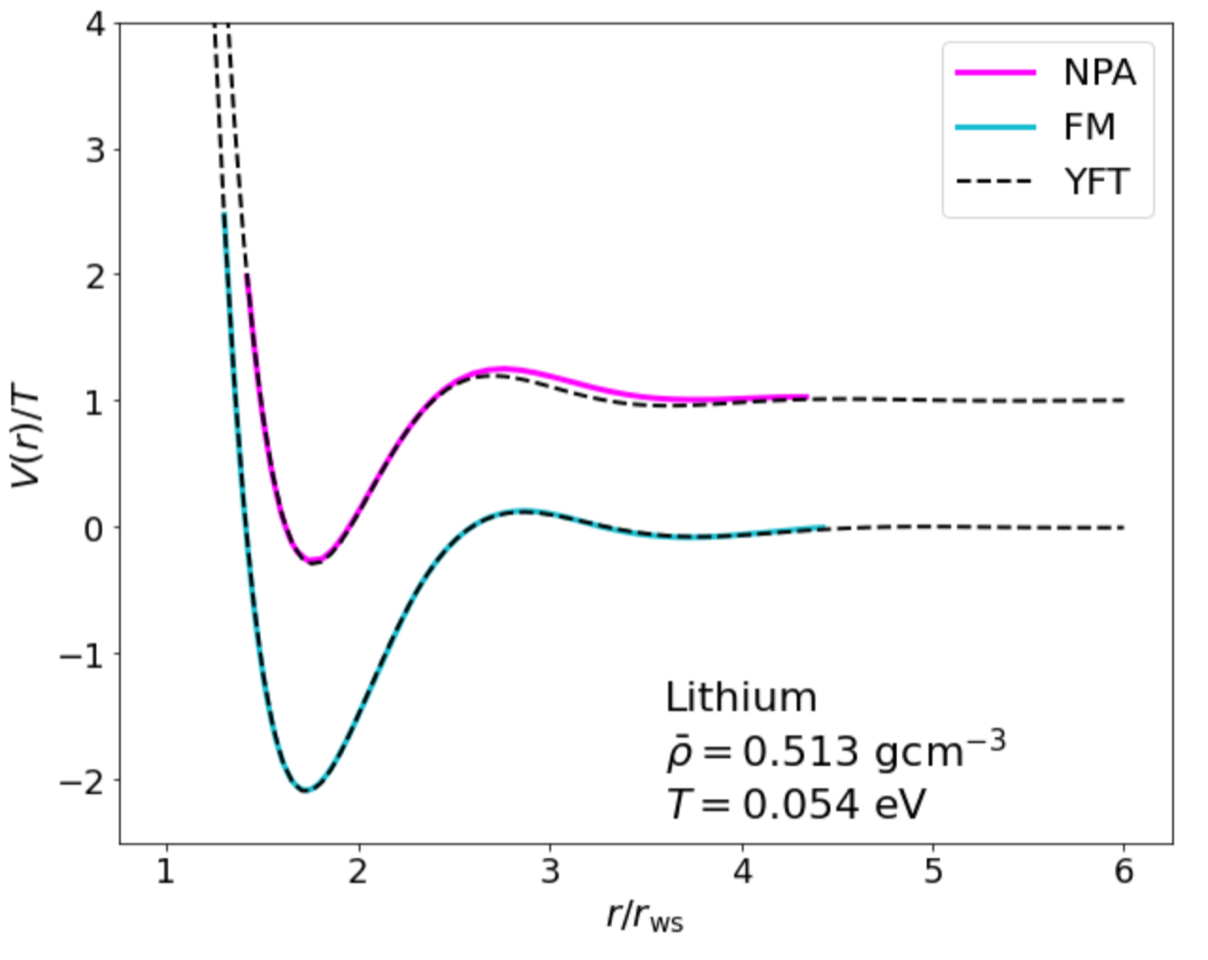}
\caption{ (Color online) A comparison of the results of fitting the YFT model,
 Eq.~(\ref{potform.eqn}), to parametrically represent the NPA and FM pair potentials
 for Li. The FM pair potential
 from QMD is extracted from QMD simulations using Eq.~(\ref{eq:LF}) \cite{Stanek21},
 while the NPA pair potential is obtained from Eq.~(\ref{pair.eqn}). The YFT model 
 correctly captures the structure of the pair potentials and allows an extension of the
FM potential to a wider range in $r$. The NPA potential and the
 corresponding YFT fit have  been shifted by unity for visual clarity. 
\label{fitcompare.fig}
}
\end{figure}

The YFT form of the potential used is taken as a sum of a Yukawa form
 $V_{\rm y}(r)$ and a thermally damped Friedel tail $V_{\rm ft}(r)$
\begin{eqnarray}
\label{potform.eqn}
V_{\rm yft} (r)&=&V_{\rm y}(r)+V_{\rm ft}(r),\\
V_{\rm y}&=&(a_{\rm y}/r)\exp(-k_{\rm y}r),\\
V_{\rm ft}&=&(a_{\rm ft}/r^3)\exp(-k_{\rm ft}r)\cos(q_{\rm ft}r+\phi_{\rm ft}).
\end{eqnarray}
Here the Yukawa form has a screening constant $k_{\rm y}$ which is of the order of
the finite-$T$ Thomas-Fermi screening length, while a different
$T$ and $ \bar{n}$ dependent screening constant $k_{\rm ft}$ defines the
 thermally damped Friedel oscillations. The Friedel oscillations 
have the usual form: a circular function phase-shifted by $\phi_{\rm ft}$
 and modulated by a $1/r^3$ decay. This six-parameter fit form is found to be
 capable of accurately reproducing the pair potentials of all the WDM systems
 containing free electrons and ions that we have studied ranging from very low temperatures
close to the melting point, to temperatures typical of the plasma state.
At higher temperatures and higher compressions, the interaction potentials become 
simpler and Yukawa-like, with the Friedel oscillations becoming increasingly
 damped~\cite{Stanek21}. Hence the potentials
discussed here focus on the more demanding regime of WDM systems, i.e., from
close to the melting temperature to about $\theta\sim 1$. In the Yukawa-like
high-$T$ high-density regimes, Thomas-Fermi models and their extensions 
like the orbital-free molecular dynamics method can increasingly
supplant $N$-atom DFT calculations. However, even in the high-$T$ range,
the NPA provides a wealth of information (via its Kohn-Sham states), ion-ion PDFs etc.,
 that are not available from Thomas-Fermi models or their extensions to
 ``orbital-free"  techniques, while being computationally very efficient.

The six parameters $a_{\rm y}, k_{\rm y}, a_{\rm ft}, k_{\rm ft},
 q_{\rm ft},$ and $\phi_{\rm ft}$ can only be approximately estimated using 
analytical models of screening. This is partly because they are also
playing the role of fitting to the form factors contained in the
actual NPA potentials that account for the finite size of the ionic 
cores.  Hence they are best obtained by fitting the NPA-derived
numerical pair potentials to the YFT fit form Eq.\eqref{potform.eqn}. The simple least-square fitting
routines available in standard open-source software or graphics packages (e.g., gnuplot or python)
are sufficient for obtaining the fit parameters of Eq.~\eqref{potform.eqn}. The fitted forms can
then be used as inputs to classical MD simulations. For example, these fitted potentials can be used
in systems with density gradients that can be described by a graded stack of layers of
uniform density and width not less than the radius of the correlation
sphere, viz., $R_c\sim 5r_{\rm ws}$. 

Furthermore, the fitted pair potentials can be compared with the
 pair potentials derived from force matching to QMD
 simulations if FM potentials are available. 
Since $N$-atom QMD calculations are carried out in a simulation
box of linear dimension $\sim r_{\rm ws}N^{1/3}$, the $N$-center potential
energy surface used in force matching is itself of restricted range, and
long-ranged oscillations are only poorly recovered in force-matched
 potentials unless a large number of particles are used. We illustrate this by
 comparing the QMD-derived C pair potential shown in Fig.~\ref{Vr.fig}(a) due
 to Whitley \textit{et al.}~\cite{whitley15} 
and the Li pair potential shown in Fig.~\ref{fitcompare.fig} due to
 Stanek \textit{et al.}~\cite{Stanek21}. 
Such QMD-derived pair potentials can be approximately extended to include their
long-ranged Friedel oscillations in two ways: (
i) they can be augmented with the Friedel tail obtained
from YFT fits to NPA pair potentials or (ii) the YFT form can be fit directly
 to the QMD data. The parametrized forms can be analytically
differentiated or manipulated more efficiently and more accurately than
numerical tabulations or ML potentials. 

\subsection{General applicability of the YFT model}
\label{generality.sec}
In the present study we explicitly provide comparisons of NPA pair potentials
and those obtained from QMD force matching. The tabulations of parametrizations given
below involve a temperature span of a factor of 10 from the lowest temperature treated.
At higher temperatures, the pair-potentials become Yukawa-like which simplifies the YFT
model accordingly by damping out the Friedel tails. 
As shown by Stanek {\textit{et al.}}~\cite{Stanek21}, the agreement between the NPA and FM
 pair potentials is usually quite good.
Furthermore, we note that the pair potentials generated via the NPA method are consistent with
 the YFT model as the NPA model used here contains an automatic check for a YFT-type
 construction that has been used for the simplifications of subsequent numerical
 applications of the potentials. Hence, all the curves marked NPA in the figures
 have YFT fits that adequately reproduce the numerically tabulated NPA or
 FM pair potentials.

It is interesting to note that many of the available AA models become
accurate mainly at the higher-$T$, higher-density regimes which are the regimes
where the pair potentials become Yukawa-like as the Friedel tails become negligible.
Similarly, the PIMC method also works best at higher temperatures
where the effects of the Friedel tails are negligible. We illustrate this in
detail using the YFT fits to WDM Al (see Sec.~\ref{Al-pot.sec}) from $T = 8$ eV to 100 eV.
Results for WDM Li and C are also given up to 100 eV and comparisons with other
methods are given when available.  

\section{Parametrized Carbon Potentials}
\label{car-pot.sec}

\begin{table*}[h]
\caption{\label{para-C.tab}The parameters of the YFT
potential $V(r)/T$ for C. The fits are for approximately the range
 $0.8<r/r_{\rm ws}<4$ and cover up to $\sim 3$ to 4 units in the energy scale
 of $V(r)/T$ as shown in Fig.~\ref{V-carb.fig}. Force-matched pair
 potentials~\cite{Stanek21} using Eq.~(\ref{eq:LF}) have been used at the
 density $\bar\rho$ = 2.267g/cm$^3$ and fit to the YFT form while the
NPA pair potentials, Eq.~(\ref{pair.eqn}), have been used for the higher-density
 YFT fits.
}
\begin{ruledtabular}
\begin{tabular}{lcccccccc}
$\bar{\rho}$ [g/cm$^3]$    & 2.267   &  2.267   &  3.5       &  3.5      &  5.0   &  5.0   & 10.0    &   10.0 \\
\hline
$T$[eV]          &   1     &   2       & 1          & 2        & 1      &  2    & 1    &   2  \\
\hline \\
$a_{\rm y}$     & 70.2535 & 43.4002  & 167.769  & 105.664 & 280.957 & 164.833 & 492.126 & 273.037 \\
$k_{\rm y}$     & 1.23545 & 0.897833 & 1.34858  & 1.42488 & 1.46306 & 1.53154 & 1.70158 & 1.75576 \\
$a_{\rm ft}$    & 60.2645 & 15.5518  & 54.2138  & 27.5023 & 23.2234 & 12.7514 & 6.04252 & 3.33661 \\
$k_{\rm ft}$    & 0.421458& 0.878615 & 0.378277 & 0.403479& 0.255881& 0.295816& 0.000006& 0.0334616 \\
$q_{\rm ft}$    & 2.23931 & 2.39948  & 2.86869 & 2.84593 & 3.20147 & 3.17996 & 3.97784 & 3.93940 \\
$\phi_{\rm ft}$          & 3.36746 & 2.75123  & 2.49232  & 2.57176 & 2.12173 & 2.22056 & 1.50360 & 1.60452 \\
\end{tabular}
\end{ruledtabular}
\end{table*}

Liquid C is usually considered to be a highly covalently
 bonded system even in the fluid state. This view is supported by ``snap-shots"
 of real-space positions of the ions obtained from QMD
 simulations. The multi-center bond-order 
potentials~\cite{Abell85,Brenner02,ghiring05} use this chemical picture to
model MCPs for C systems. However, as is evident from DFT,
only the long-time average is relevant to determining the
 thermodynamics and static structure factor since the
effects of the transient bonds average out in $l$-C~\cite{DWP-carb90}.
In QMD, this average is explicitly carried out using millions of
 configurations of solid-like ionic clusters, without taking
 advantage of the spherical symmetry of the fluid or the possibility
of using an ion-ion XC-functional to treat many-center effects.  
While the chemical-bonding picture does not use the Fermi-liquid
character of $l$-C, the NPA approach exploits it to the
fullest extent in the YFT model. 

\begin{figure}[t]
\includegraphics[width=.95\columnwidth]{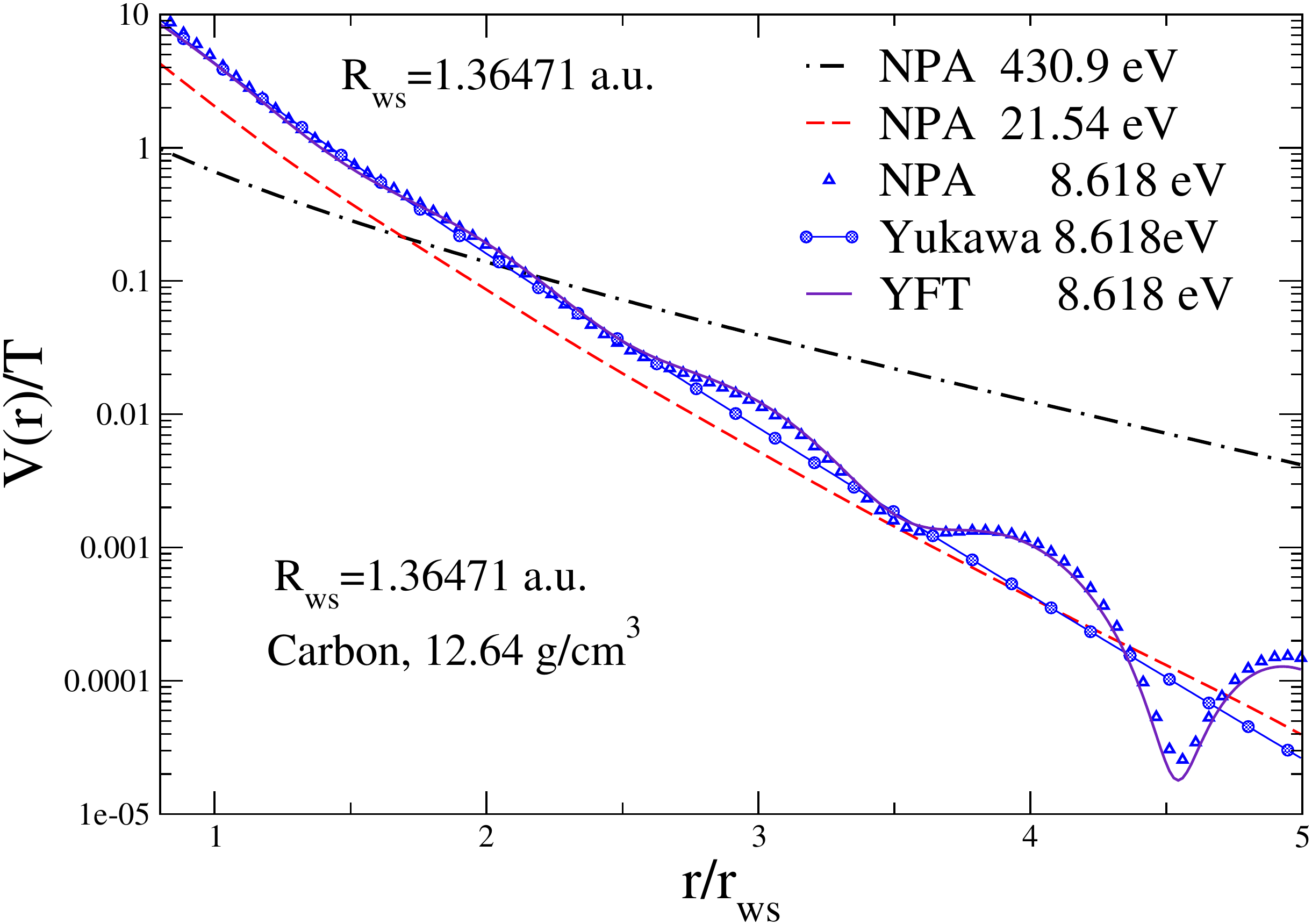}
\caption{(Color online) The NPA C-C pair potentials for WDM C studied
by Driver {\textit{et al.}}.~\cite{driver12}. The $y$-axis ($V(r)/T$) is logarithmic
and the quasi-linearity of the plots for $V(r)$ at $T=430.9$ eV and 21.54 eV
establish the Yukawa-like character of the pair-potentials. At $T=8.618$ eV, the pair potential
 contains Friedel oscillations for
 $r > 1.5r_{ws}$; the NPA potential is well fitted by the YFT form.
\label{pimc-pot.fig}
}
\end{figure}

\begin{figure}[t]
\includegraphics[width=.95\columnwidth]{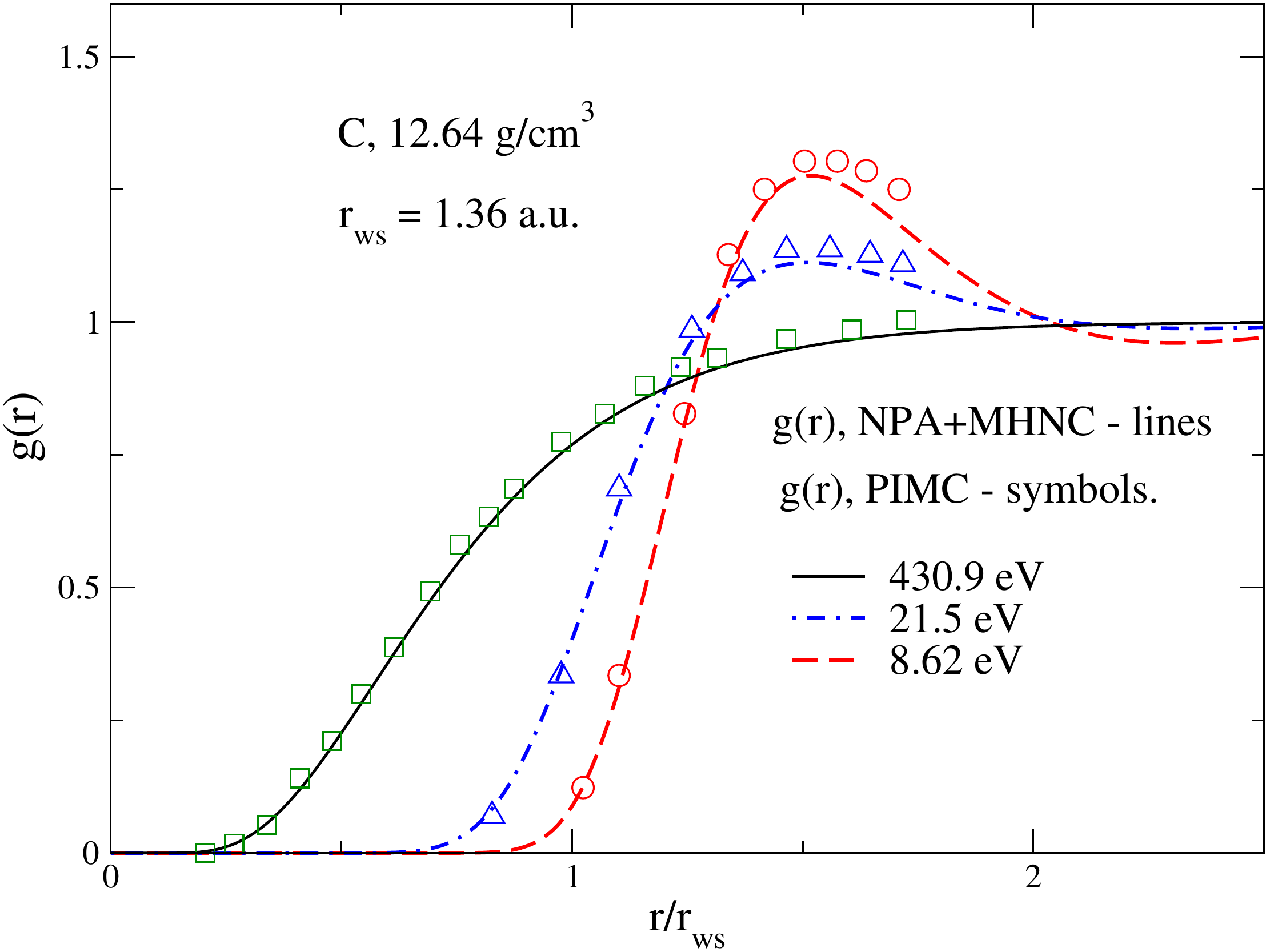}
\caption{(Color online) The C-C PDFs $g(r)$ from the
NPA potentials, from PIMC and QMD of Driver {\textit{et al.}}~\cite{driver12}.
\label{pimc-gr.fig}}
\end{figure}

We calculate the NPA pair potentials and the $g(r)$ of $l$-C, at
density 12.64 g/cm$^3$,  at $T = 8.62 $ eV
(1x$10^5$K), 21.5 eV (2.5x10$^5$K), and 430.886 eV (5x10$^6$K) to compare
with the PIMC and QMD  calculations of Driver and
Militzer~\cite{driver12}. The mean ionization $\bar{Z}$ obtained from
the NPA changes from $\bar{Z} = 4.0$ at low $T$, to $\bar{Z} = 4.05$ by $T=100$ eV, and
 reaches $\bar{Z} = 5.758$ by $T = 430.9$ eV -- the highest $T$ used by Driver and
Militzer~\cite{driver12}.

The YFT fits to the NPA potentials that reproduce the
PIMC $g(r)$ data even at low-$T$ and at lower density
are given in Fig.~\ref{pimc-pot.fig}. The comparison of the
NPA $g(r)$ obtained from the NPA potentials with the $g(r)$ from PIMC
are given in Fig.~\ref{pimc-gr.fig}. The NPA $g(r)$ are obtained from
the pair potentials using the MHNC equation. This is an inexpensive alternative
 to carrying out an MD simulation. We caution that if using MHNC in place of MD,
 one needs to specify an accurate bridge function in the MHNC equations. 
Additionally, although hard-sphere approximations to the bridge function
 are often employed in the MHNC equations
by appealing to a universality principle~\cite{Rosenfeld93}, it
should be used with some care as this model may not be valid for 
certain systems. For example, the bridge function can be set to zero for
C, Si, and Ge in the liquid state at normal densities, and near their
melting points when their structures are controlled mainly by the
scattering of electrons by ions from one edge of the Fermi surface to
the opposite edge, giving rise to $2k_F$ scattering effects~\cite{cdw-carb22}
that control the ionic structure, rather than packing effects.

\begin{figure}[t]
\includegraphics[width=.95\columnwidth]{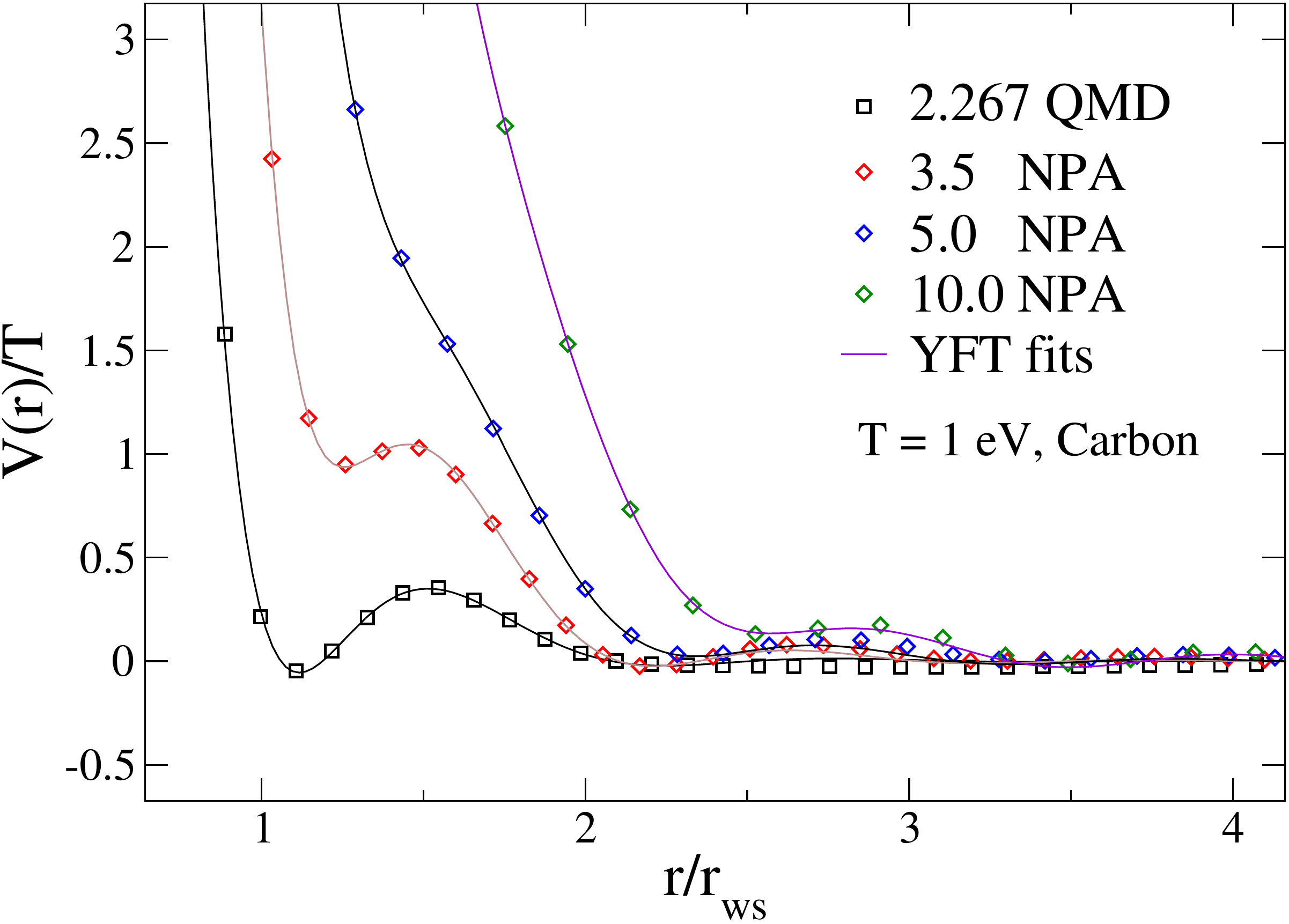}
\caption{ (Color online) The C-C pair potentials generated from FM and NPA
(denoted by shapes) and their parametrized forms (continuous lines)
based on the YFT model. The curves are identified by the numerical
value of the corresponding systems density. The curve 2.267 QMD is the
C-C pair potential obtained by FM to QMD
 data using Eq.~(\ref{eq:LF}) for
 $\bar{\rho}$ = 2.267 g/cm$^3$
at $T=1$ eV~\cite{Stanek21}. The pair potentials at higher densities
were generated using the NPA, Eq.~(\ref{pair.eqn}),
as indicated in the legends.
\label{V-carb.fig}
}
\end{figure}

In Table~\ref{para-C.tab}, we present a sample of parametrizations for the YFT  potentials for
 C along the $T=1$ and 2 eV isotherms, for the range of
 densities $\bar{\rho} = 2.267$ g/cm$^3$ up to 10 g/cm$^3$. At higher $\theta$ and
higher densities, the potentials increasingly approximate a Yukawa-like form. The
linear response potentials are applicable for $\bar{\rho}\ge$ 3 g/cm$^3$ 
 where they recover the structure factor and pressure obtained
via QMD simulations~\cite{cdw-carb22}. The low-density range from
 $\bar{\rho}<3$ g/cm$^3$ and extending towards the graphite density and
below, cannot be accurately computed using the linear-response model given by Eq.~\eqref{uei.eqn}.
Therefore, we use a FM pair potential to construct a 
parameterization with the YFT model and provide parameters for the
density $\bar{\rho}=2.267$ g/cm$^3$. We note that the mean ionization of C is such that
 $\bar{Z}=4$ for the densities and temperatures studied here. 

Higher densities and temperatures (e.g., 10-100 times the diamond
 density and in  the 10$^6$ K temperature range) have been studied using the
 NPA~\cite{cdw-Carbon10E6-21}, and the pair potentials under such conditions
 can be accurately fitted to the YFT form.
To illustrate the fact that the YFT potentials reduce to simple Yukawa-like
potentials for sufficiently high-$T$, we give the parametrizations for the
Yukawa-like pair potentials of $l$-C at the graphite density in Table~\ref{Ypara-C.tab}
 of Appendix \ref{sec:tabulations}.

The parameters given in Table~\ref{para-C.tab} 
reflect the character of a screened metallic ion. This is confirmed by the
value of the screening constant $k_{\rm y}$ which is quite close to the
Thomas-Fermi screening wavevector, while the wavevector $q_{\rm ft}$
 associated with 
the Friedel oscillations, is approximately 2$k_F$; the Friedel oscillations
decay as $1/r^3$
as expected for a metallic fluid. The screening constant $k_{\rm ft}$ introduces
a damping of the Friedel oscillations. For systems such that  $\theta$ is small, this
damping constant may be set to zero and the Friedel oscillations are similar to
those familiar in metal physics. Since the $l$-C systems studied
here are such that $\theta$ is very small, we could have set $k_{\rm ft}=0$ and used
a simpler five-parameter fit which is equally capable of fitting to the QMD or NPA
data. We will make this simplification for treating $l$-Al as is done
 in Sec.~\ref{sec:Al} and presented in Table~\ref{para-Al.tab}.

\section{\label{sec:Al}Parametrized Aluminum Potentials}
\label{Al-pot.sec}
We study normal density Al from $T=0.5$ eV to 100 eV, and in the density
 range $\bar\rho = $ 2.0 g/cm$^3$ to 6 g/cm$^3$ at
 $T = 1$ eV. The normal solid-state density $ \bar\rho$ is 2.7 g/cm$^3$,
 while the isobaric $\bar\rho$ is 2.375 g/cm$^3$
 at its melting point ($T =$ 933.47 K).
 Unlike the C ion with $\bar{Z}=4$ which is almost point-like -- having only
 the 1$s$ of bound electrons -- Al$^{3+}$ has a large core and its form factor is
 an important feature of its pseudopotential. Nevertheless, we find that the YFT
 model works satisfactorily for $l$-Al as well because the
YFT parameter set is flexible enough for doing ``double duty". What we mean by this is that
 the parameters of the YFT model have more fitting capacity than would be found
 in a strictly defined physical model. Such a strictly defined model would require a parametrization
of the form factor of the pseudopotential and a parametrization of the electron
response function, so that Eq.~\ref{pair.eqn} is itself fully parametrized.

A highly non-local pseudopotential rather than the simple $s$-wave form has been
used in linear-response~\cite{DRT75} and in QMD codes like the VASP~\cite{VASP} and ABINIT~\cite{ABINIT}
although we find this to be unnecessary in NPA applications to uniform-density
systems and cubic solids. Liquid Al differs from  $l$-C in that the
 main peak of the structure factor is not located at 2$k_F$ as in C, Si,
 and Ge~\cite{cdw-carb22} at their typical densities. The YFT parameters
 that fit the NPA potentials for $l-$Al are given in Table~\ref{para-Al.tab} of
 Appendix~\ref{sec:tabulations}.
The WDM form of $l$-Al very close to its melting point, and under more extreme
 conditions have been studied using the NPA and reported in previous
 publications~\cite{eos95,cond3-17,PDWBenage02}.

\begin{figure}[h]
\includegraphics[width=.95\columnwidth]{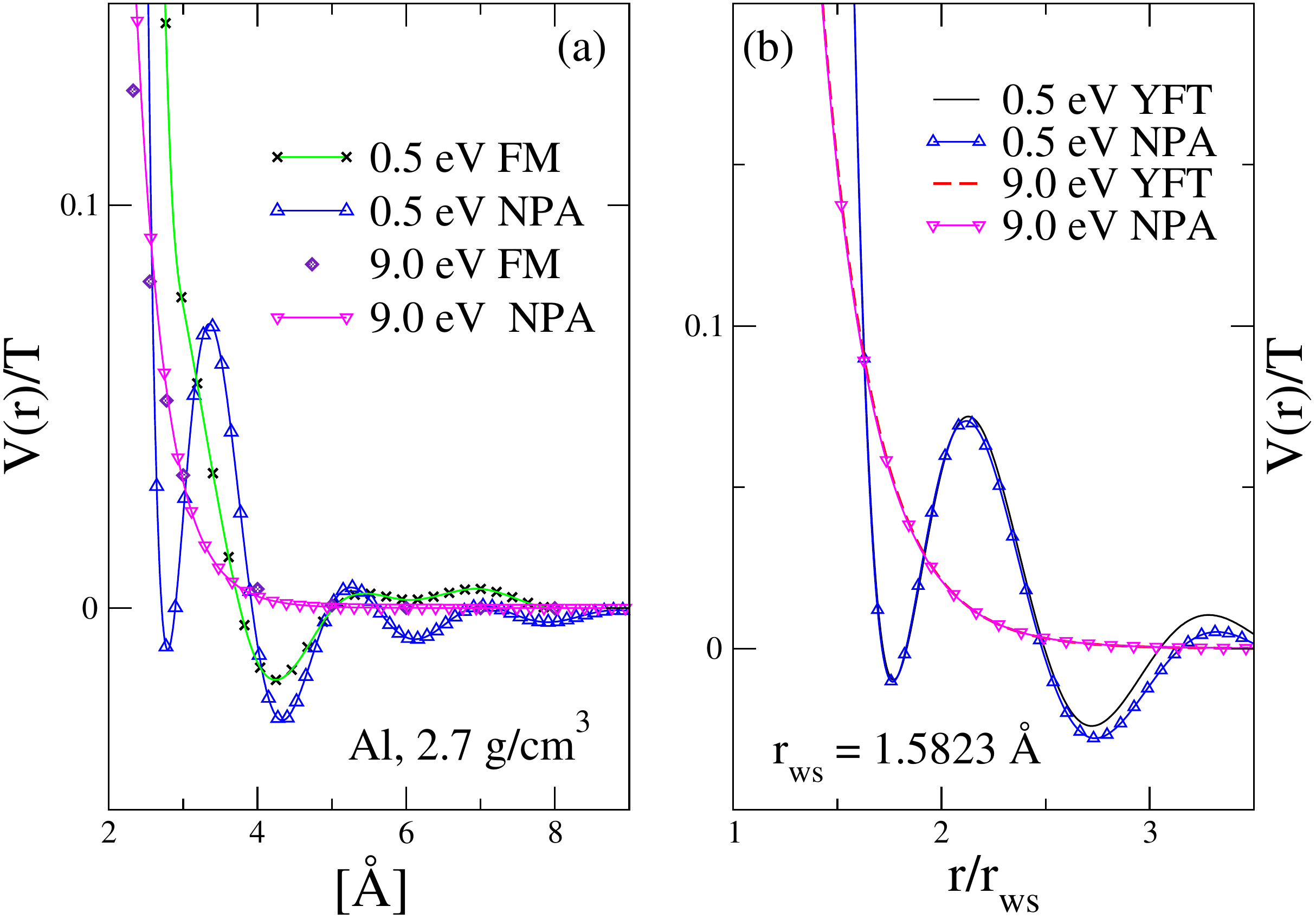}
\caption{ (Color online) (a) The Al-Al pair potential at $\bar\rho$ = 2.7 g/cm$^3$
from NPA,  Eq.~(\ref{pair.eqn}), and from 
  FM using Eq.~(\ref{eq:LF}) on QMD data~\cite{Stanek21}, for temperatures
  $T = $ 0.5 eV and 9 eV. The oscillatory terms seen in the NPA potential,
 larger at the melting point ($T = $ 0.083 eV), are already less than a tenth
 of the thermal energy by $T = $ 0.5 eV, and the FM pair potentials recover
 an average value.
 (b) The NPA potentials and the YFT fits in the region relevant to
the main peak-structure of $g(r)$. The YFT fit reduces to a Yukawa form
at $T = $ 9 eV \footnote{The YFT
parameters are available from the authors.}.}
\label{Vr-Al.fig}
\end{figure}

The Friedel oscillations in $V(r)$ are of great importance
 for the potential near the melting point of Al. Under these conditions,
 the first peak of the $g(r)$ locates at a positive energy turning point of a Friedel
 oscillation~\cite{DRT75,cdw-aers83,HarbEOSPhn17,DSF18}.  
The NPA potentials between $T = $0.5 eV and 9.0 eV have been compared with FM
potentials obtained using Eq.~(\ref{eq:LF}) from QMD simulations~\cite{Stanek21}. In Fig.~\ref{Vr-Al.fig}
we show two cases of Al at $\bar{\rho} = 2.7$ gcm$^{-3}$, viz., one for $T = $ 0.5 eV and
 another for $T = $ 9.0 eV. The $T = $ 0.5 eV case is typical of the YFT regime where
 Friedel oscillations are non-negligible, whereas the $T = $ 9.0 eV case typifies the
 regime where the Friedel oscillations have become sufficiently damped and the YFT
 potential is approximately Yukawa-like.
 
At low-$T$, the NPA potentials have oscillatory tails while the FM potentials
 capture them only partly, as already seen in the C potentials in Fig.~\ref{Vr.fig}(a).
 However, the FM pair potentials and the more oscillatory NPA potentials nevertheless yield PDFs
 and ion-ion structure factors that are in close agreement. By $T = 9.0$ eV, which corresponds to
 $\theta=0.77$, the potentials are no longer oscillatory and only the
  Yukawa form remains. This is in fact often the case by $\theta\gtrsim 2/3$ even at other densities,
 and for many elements. As $T$ increases, $\theta$ increases and the Yukawa approximation
 improves, but this is modified by changes in the mean ionization $\bar{Z}$ which increases
$E_F$, thereby countering the accuracy of the Yukawa form. Nevertheless $\theta$ steadily
 increases in WDM Al from $\theta=1.549$ for T = $20$ eV where $\bar{Z}$ = 3.495 to $\theta=4.567$,
 for T = 100 eV where $\bar{Z}$=7.721. Thus the Yukawa form continues to be a reasonable
approximation in spite of increased ionization; the values of the parameters for the
range $T = 8$ eV to 100 eV for normal density Al is given in Table~\ref{Ypara-Al.tab} of
 Appendix~\ref{sec:tabulations}.

Although the potentials have become Yukawa-like, the ion-ion $g(r)$ and $S(k)$
are still strongly coupled as the plasma parameter $\Gamma=\bar{Z}^2/(r_{\rm ws}T)$
 exceeds unity,  being $\Gamma = 8.35$ at $T = 10$ eV and $\Gamma = 6$  at $T = 80$ eV. Hence $g(r)$ must
 be computed using MHNC or with MD. A simple but useful approximation may
be obtained by a one-component-plasma analogy based on the value of
 $\Gamma$~\cite{Clerouin14Gamma}. The fact that the potentials reduce
 to Yukawa-like forms in at least the $\theta\gtrsim 1$  regime opens the door for a
 semi-analytic theory of  these WDM systems for a wide range of conditions.

In fitting the NPA potentials to the YFT form we have simply used no thermal damping
on the Friedel term ,i.e., setting $k_{\rm ft}$ equal to zero. This shows
that the parameters of the fit form ``do double duty'' and their values are not
necessarily related to the values expected from basic theoretical models, although quite
close to them. This fortunate flexibility is partly associated with
 the freedom in the selection of the range of $r$ used in the fits. Standard
pseudopotential formulations also have such flexibility arising from how
the atomic core is excluded in the pseudopotential.
 In general, the fit range is selected to be that of the
 ion-ion $g(r)$, viz $r_{\rm ws} \le r < R_c$.

\begin{figure}[h]
\includegraphics[width=.95\columnwidth]{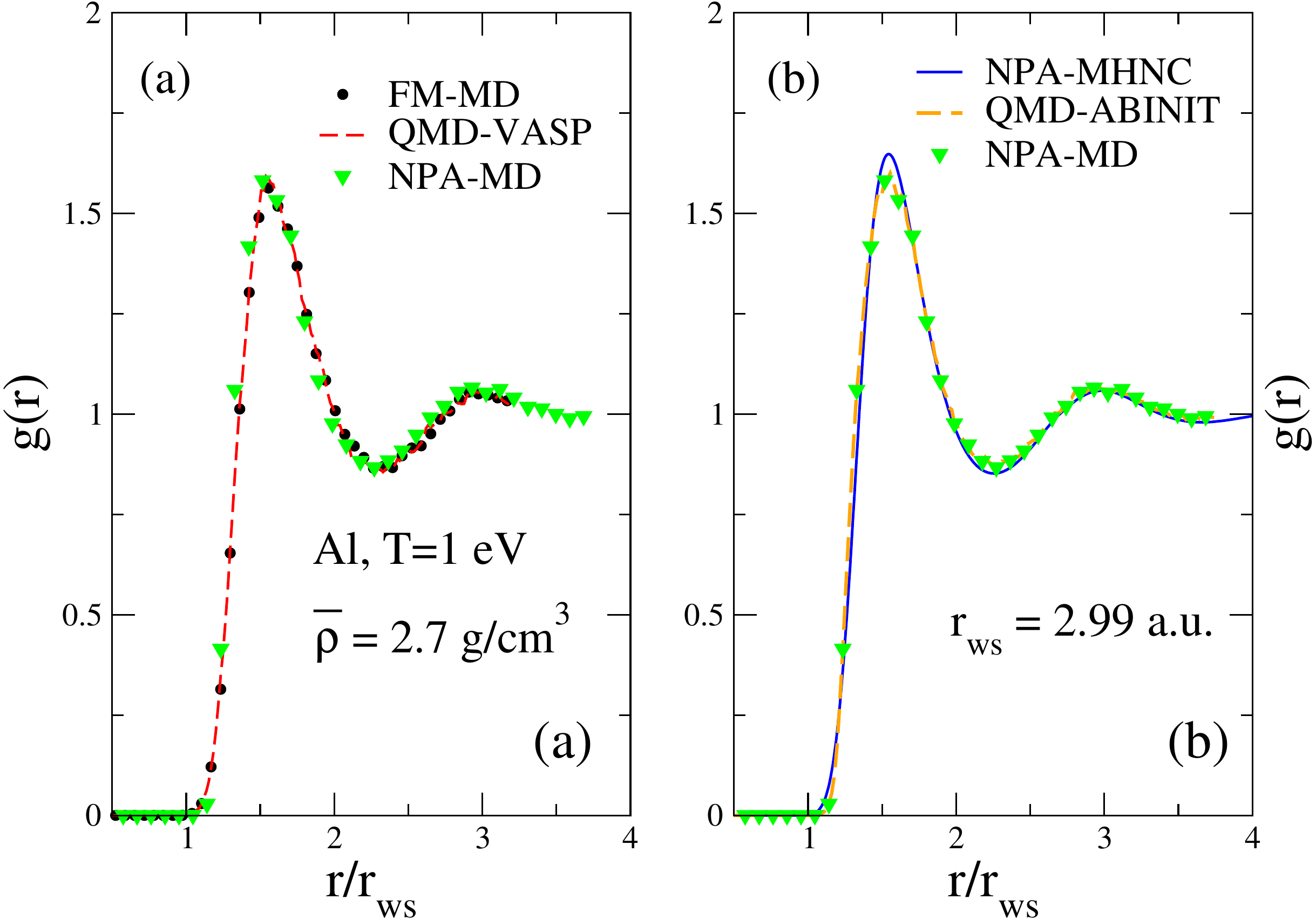}
\caption{ (Color online) (a) The PDF $g(r)$ of Al at $T = $ 1 eV and $\bar\rho = 
$ 2.7g/cm$^3$. The PDFs are generated using QMD~\cite{Stanek21} \textit{and} MD with the
FM and NPA pair potentials.
(b) The PDF from QMD~\cite{DSF18} and with the NPA pair potential, Eq.~(\ref{pair.eqn}). 
The NPA PDFs are generated from MD and also with MHNC. The
 NPA-MD and the FM-MD agree closely although the NPA potential has more oscillatory
 structure as seen in Fig.~\ref{Vr-Al.fig}. 
\label{gr-Al.fig}
}
\end{figure}

It should be noted that the pair potential defined by Eq.~(\ref{pair.eqn}) does
 not contain interactions with the polarizable atomic cores of the two interacting
Al ions, or van der Waals interactions. The YFT fit-forms also do not specifically
 contain  such interacting terms, although the parameters have some flexibility to
 absorb such contributions approximately. 
 How these concerns could be included within a purely DFT scheme were discussed in the Appendix
 of  Ref.~\cite{eos95}. These concerns are also important in systems that contain a
significant mixture of neutrals, e.g., Ar plasmas at $T \sim$ 1$-$2 eV and
normal density~\cite{Stanek21} where an argon mixture is treated using g the NPA.
 At the temperatures and densities considered here, 
the dominant interactions are those included in Eq.~(\ref{pair.eqn}), and hence
we will not discuss core-polarization effects further. 
 
In Fig.~\ref{gr-Al.fig}(a) we compare the $g(r)$ generated directly within QMD,
 and labeled QMD-VASP
with the $g(r)$ obtained from MD using the FM potential extracted from the QMD
using Eq.~(\ref{eq:LF}).
In Fig.~\ref{gr-Al.fig}(b) we compare the $g(r)$ generated directly from QMD using ABINIT,
 with the PDFs
obtained from the NPA pair-potential using MD, and using MHNC~\cite{DSF18}, showing that
 the MHNC yields $g(r)$s of good accuracy. 
 We may also note that the NPA pair potential provides an $S(k)$ for Al that
 not only agrees with the
static $S(k)$ obtained by QMD simulations, but also accurately renders the dynamic ion-ion
 structure factor $S(k,\omega)$ obtained from DFT-MD microscopic simulations
 available at $T = $ 1 eV and at $\bar{\rho}$ = 2.7 g/cm$^3$~\cite{DSF18}.

\section{Parametrized Lithium Potentials}
\label{li-pot.sec}
Lithium in its WDM state has not been as extensively studied as Al or C.
 However, accurate data for the structure factor $S(k)$ of
 $l$-Li near its melting point as well as attempts at theoretical predication
 are available~\cite{Salmon2004}. Since theories should preferably be
 tested against actual experimental data  rather than against simulation
results, we first study this case, i.e., $l$-Li at $\bar{\rho}$ = 0.513(4) g/cm$^3$ and
at a temperature of $T =$ 197$^{\circ}$C (which corresponds to $T = 0.0405$ eV
in the units used here). At this very low temperature and nearly the solid-state
density, it is easy to imagine that pair potentials would be totally
 inadequate and that a multi-center approach would be needed.
However, our results show that the NPA pair potential, taken together
with the MHNC integral equation where the bridge term is modeled by a
hard-sphere fluid very successfully predicts the experimental data. The
packing fraction $\eta=0.3$ of the hard-sphere fluid is determined using the 
Lado-Foiles-Ashcroft free-energy minimization criterion~\cite{LFA83}.

\begin{figure}[t]
\includegraphics[width=.95\columnwidth]{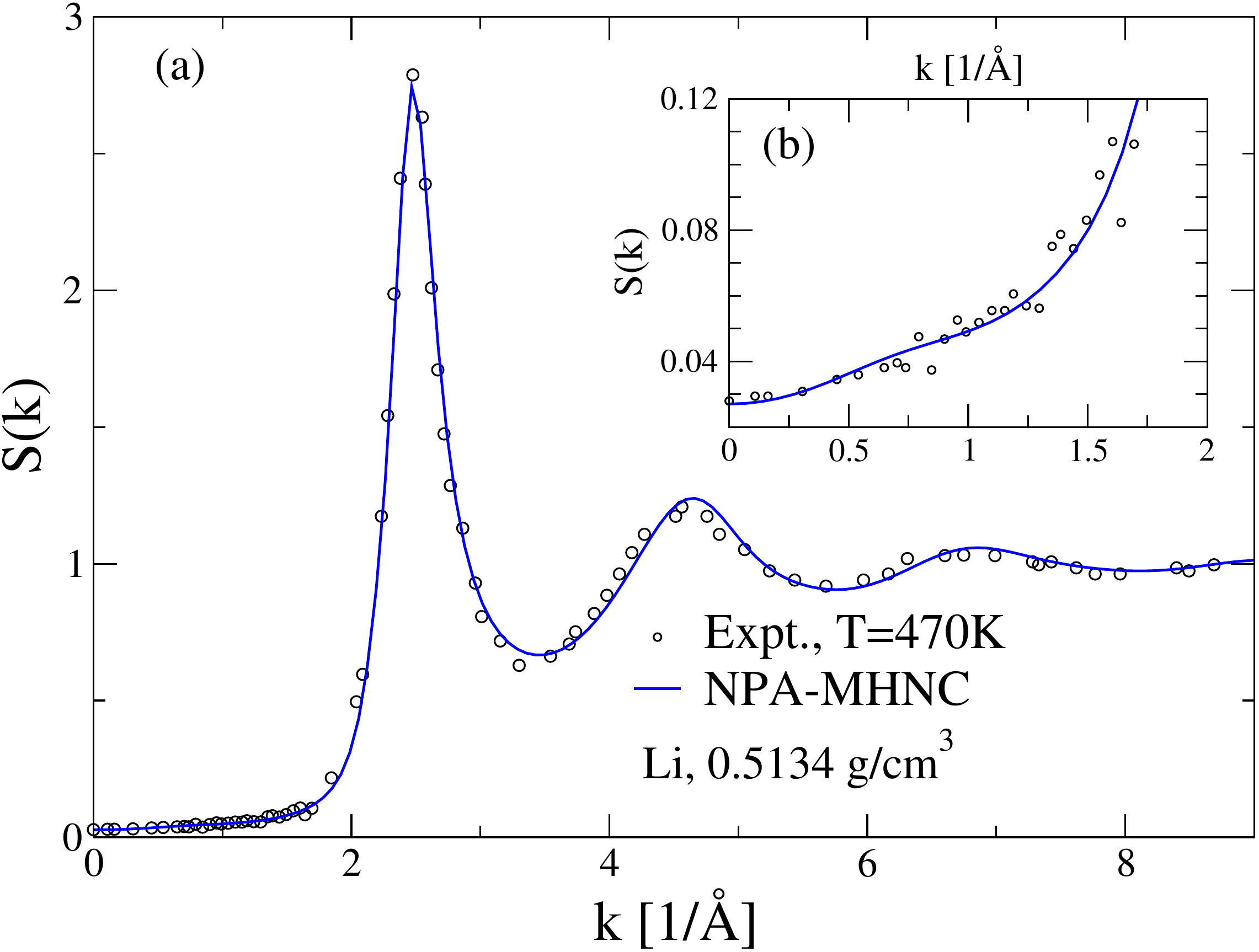}
\caption{ (Color online) (a) The neutron-scattering structure factor
 of Li at $T = $ 470 K is displayed as circles, $S(k)$~\cite{Salmon2004}.
 It is compared with the theoretical $S(k)$ from  the NPA-MHNC calculation.
 The bridge contribution is that of a hard-sphere fluid with a packing
 fraction of $\eta=0.3$, given by the Lado-Foils-Ashcroft criterion.
 The inset (b) shows the low-$k$ behavior and the accurate recovery
of the $k\to 0$ limit required to satisfy the compressibility sum rule. 
The Li-Li NPA pair potential for this case, $\bar{\rho} = 0.5134$ g/cm$^3$ is
displayed in Fig.~\ref{Li-Vr.fig}.\label{Li-Salo.fig}
}
\end{figure}

Stanek \textit{et al.}~\cite{Stanek21} have compared the NPA potential and
 its predictions for Li  at $\bar\rho = 0.513$ g/cm$^{3}$ and $T=  0.054$ eV,
 with those from FM potentials obtained from QMD simulations using
 Eq.~(\ref{eq:LF}), obtaining excellent  agreement. As the Fermi energy
 of Li at this density is $E_F \sim$ 5 eV, with $\theta \simeq 0.011$, the
electron subsystem is effectively highly degenerate. The NPA potentials were
 found to work very well even for solid-density Li as the phonons calculated
 using the NPA pair potentials were in good agreement with experimental
 phonon dispersion data~\cite{HarbEOSPhn17}.
 It has also been verified that the NPA pair potential for Li at $T = 0.173$ eV
 generates PDFs in good agreement with the QMD data at
 $\bar{\rho}=0.85$ g/cm$^3$~\cite{Kietzman08}.

\begin{figure}[t]
\includegraphics[width=.95\columnwidth]{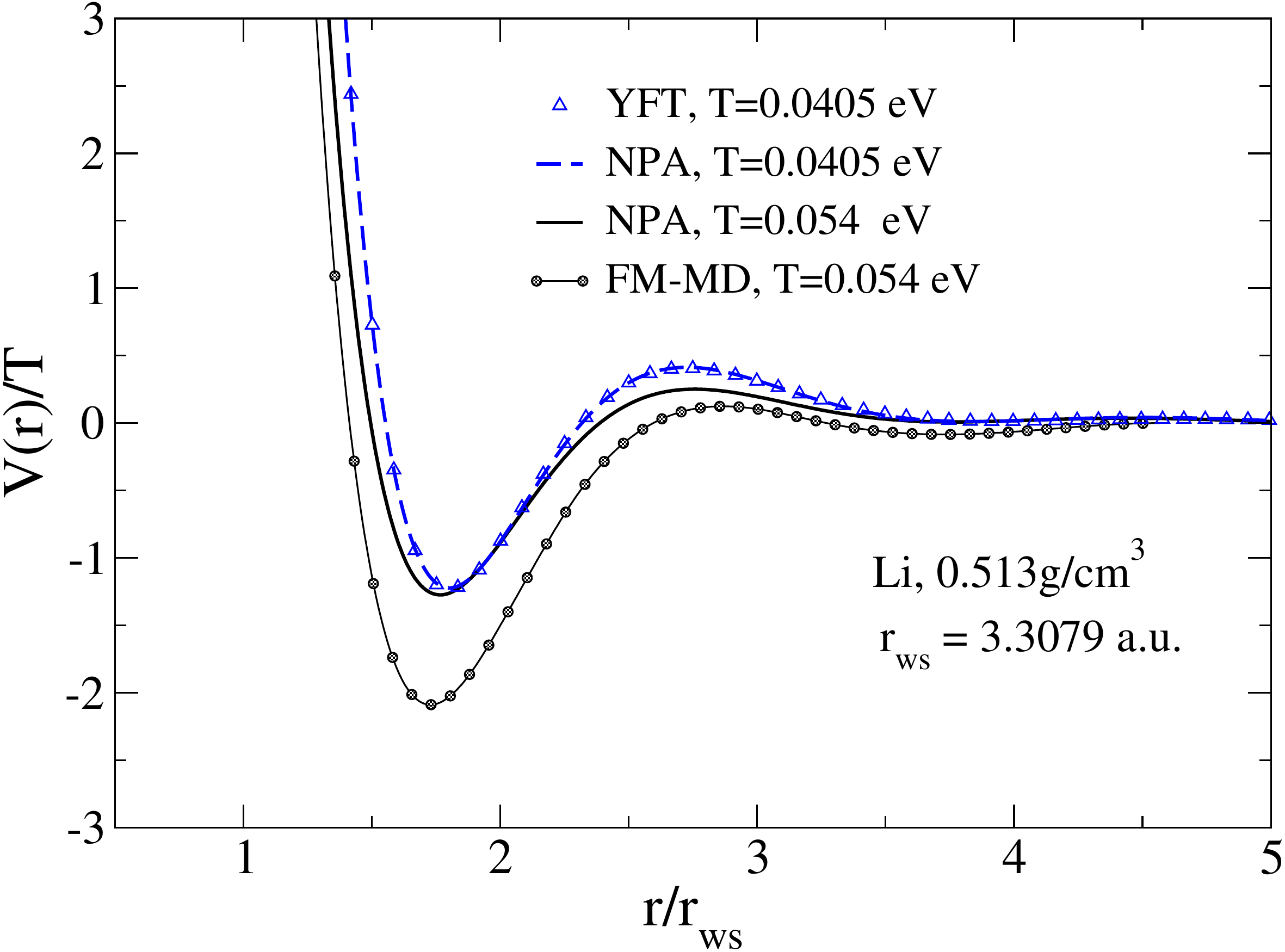}
\caption{ (Color online) The Li-Li pair potential. The FM pair potential obtained by
 Stanek \textit{et al.}~\cite{Stanek21} (denoted as QMD-FM) corresponds to Li at
 $\bar{\rho} = 0.513 $ g/cm$^{3}$ and $T$ = 0.054 eV; the corresponding NPA pair potential
calculated from Eq.~(\ref{pair.eqn}) is given by solid lines.
The NPA pair potential at $T = 0.0405$ eV = 470 K is denoted by a dashed line and
 corresponds to a case for which
 neutron-scattering $S(k)$ data are available (see Fig.~\ref{Li-Salo.fig}).
\label{Li-Vr.fig}
}
\end{figure}

While the above cases are for high-electron degeneracy, Harbour
 \textit{et al.}~\cite{cond3-17} found good agreement between QMD calculations of
 various aspects of X-ray Thomson scattering  and those
from NPA methods for the density $\bar\rho = 0.6$ g/cm$^{3}$ and $T=4.5$ eV.
 In this case, the electron system is at a temperature close to the Fermi energy
 and is partially classical in its behaviour. Hence we may
 conclude that currently available QMD data confirm the validity of the NPA potentials
 for WDM-Li from the solid state to the plasma state. A comparison of the FM Li-Li pair
 potential using Eq.~(\ref{eq:LF}) at $T = 0.54$ eV and $\bar{\rho} = 0.513$ g/cm$^{3}$
 and the corresponding NPA Li-Li potential is given in Fig.~\ref{Li-Vr.fig}.

\begin{figure}[t]
\includegraphics[angle=0, width=0.95\columnwidth]{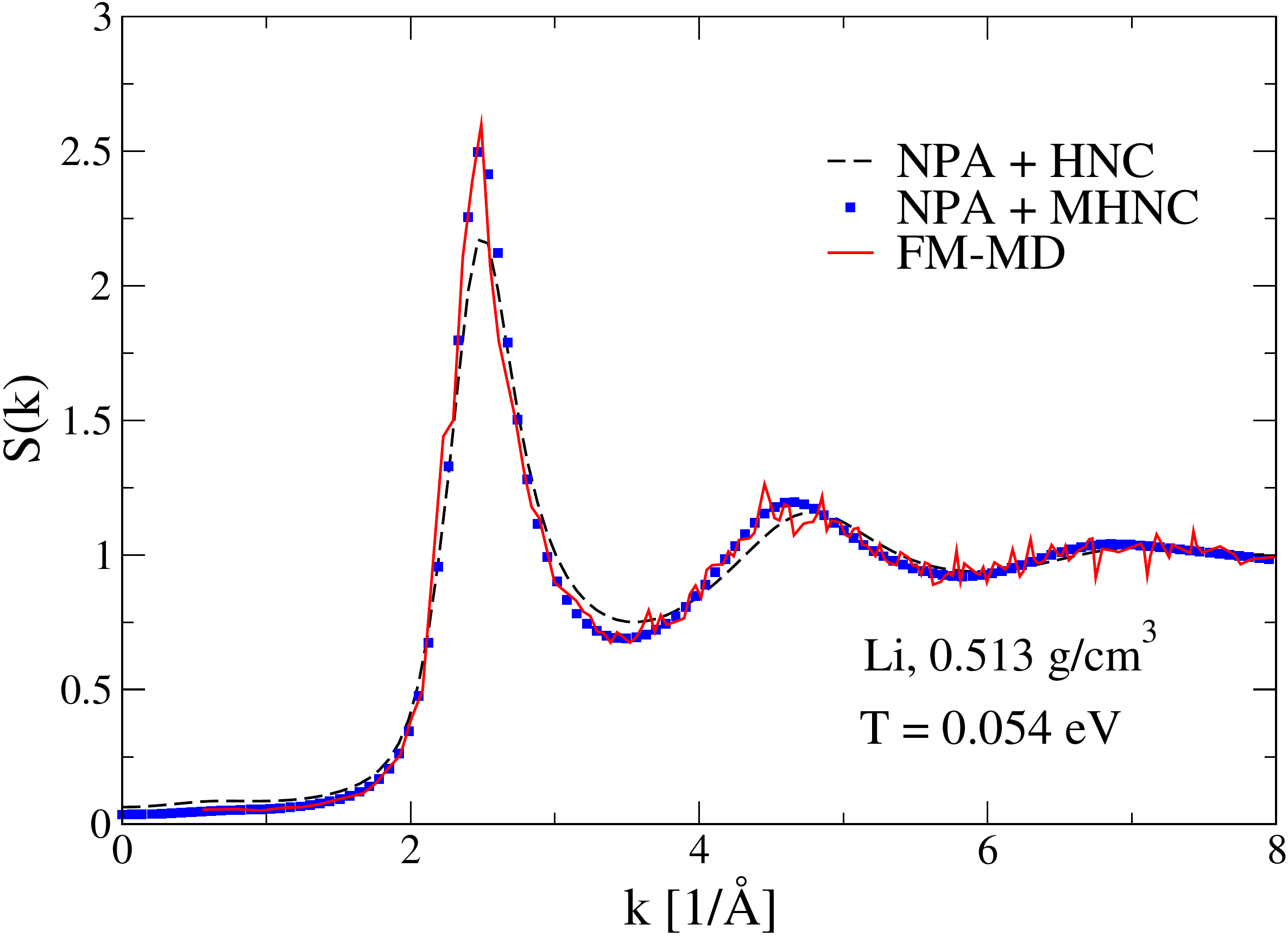}
\caption{ (Color online) The $S(k)$ from the NPA pair potential generated
 using HNC and MHNC, and from the FM pair potential,
Eq.~(\ref{eq:LF}), using MD are
 displayed.
\label{Li-sk-Luke.fig}
}
\end{figure}

\begin{figure}[t]
\includegraphics[angle=-0, width=0.95\columnwidth]{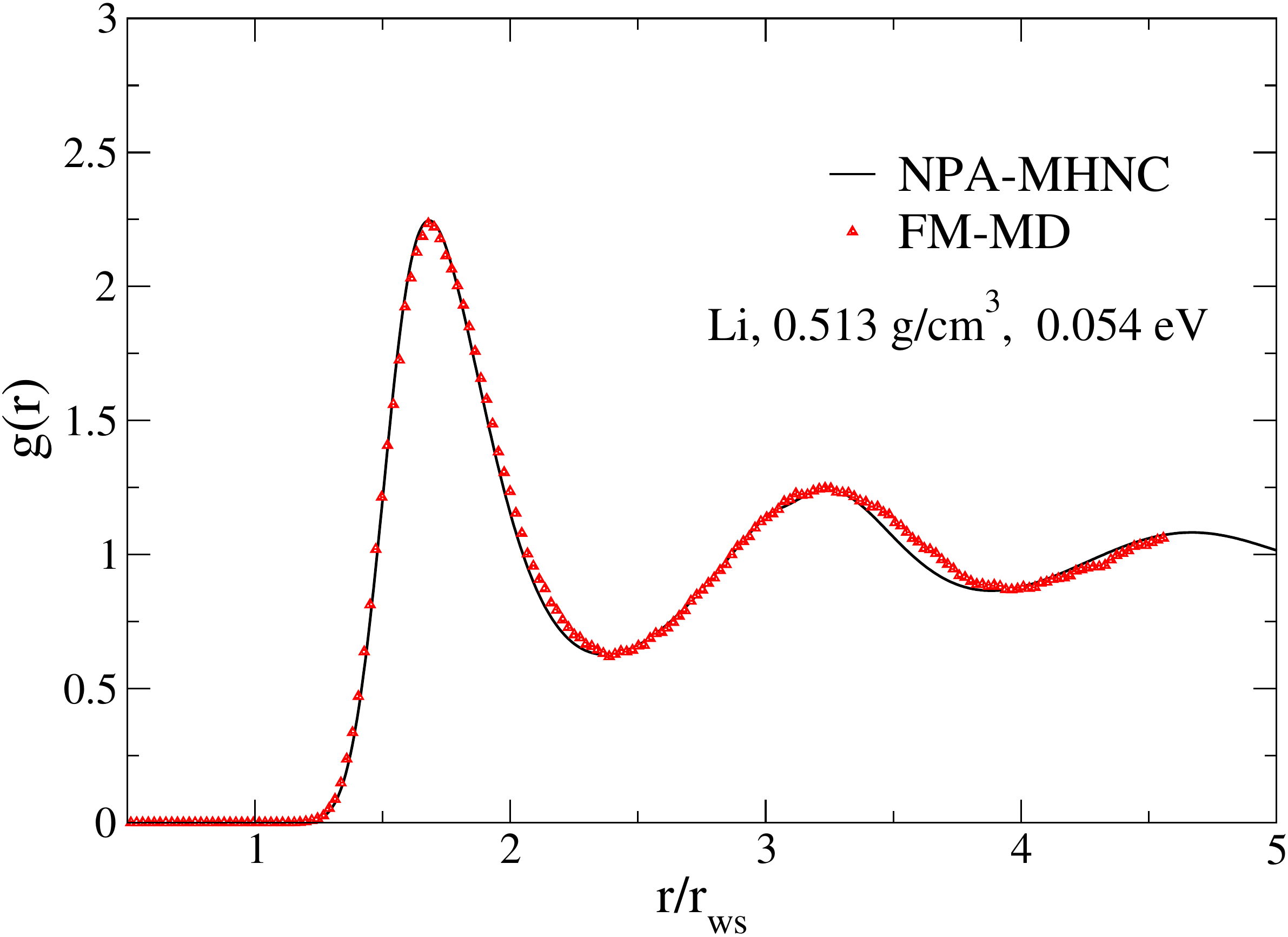}
\caption{ (Color online) The PDF $g(r)$ from the Li NPA pair potential generated
 using MHNC, and from the FM QMD pair potential~\cite{Stanek21}, Eq.~(\ref{eq:LF}), using MD are displayed.
\label{Li-gr-Luke.fig}
}
\end{figure}

Unlike in the case of Al, the difference between the FM $V(r)$
and the NPA $V(r)$ at their minima is more significant compared to the 
case of Al at $T = $ 1 eV (Fig.~\ref{Vr-Al.fig}), being of the order of the
thermal energy of the system at $T = 0.054$ eV. However, this difference
is not reflected in the predicted properties, e.g., the $S(k)$ or $g(r)$ Li, as
seen in Fig.~\ref{Li-sk-Luke.fig} and Fig.~\ref{Li-gr-Luke.fig} respectively.
In Fig.~\ref{Li-sk-Luke.fig} we show the $S(k)$ obtained from the HNC, i.e., MHNC
 with the bridge function set to zero as well
 as the MHNC with the proper LFA-optimized bridge term~\cite{LFA83}, displaying the role of the
 bridge contribution to the ionic structure.

In Table~\ref{para-Li.tab} of Appendix~\ref{sec:tabulations} we provide YFT fits to Li-Li pair
potentials generated using the NPA electron density and the linear-response pair
 potential given by Eq.~(\ref{pair.eqn}). 
In most cases, we have set the parameter $k_{\rm ft}$
 controlling exponential damping term of the Friedel oscillations to
zero, except in the case of the density $\bar\rho =$ 0.85 g/cm$^3$ where we have
 included it as a parameter, merely to illustrate the fact that there is
 significant interplay among parameters, and consequently, a numerical fit of
 similar quality  can usually  be obtained even if $k_{\rm ft}$ is set to zero.
We give YFT fits for several densities
 from $\bar\rho = 0.513$ g/cm$^3$ to 1 g/cm$^3$, and at temperatures $T = 
0.054$ eV and 1.0 eV in all cases. We also give the YFT fit at $T=4.5$ eV and
 $\bar\rho =$ 0.513 g/cm$^3$ which corresponds to $\theta=0.983$ where the
fit form reduces to a Yukawa form as the Friedel oscillatory term is found to be
 negligible. Just as in Al, we have a regime $\theta\gtrsim 1$ where the
pair-potentials of the Li WDM system become Yukawa-like in spite
of increasing ionization with increasing temperature. The Yukawa-like parameters
that fit the NPA potentials in this regime are given in Table~\ref{Ypara-Li.tab} of
the Appendix~\ref{sec:tabulations}.

\section{discussion}
\label{disc.sec}
There are multiple motivations for constructing parametrizations of pair potentials based on an
 approximate physical model with very few parameters rather than using a highly
 flexible purely numerical approach fitted to QMD data. First,
the parametrization provides some validation to the associated physical picture. Second, the fits
can be used in simulation codes which call for repeated evaluations of the potentials
and total energies  as a function of
position, temperature, and density. Third, the model potentials can be used
to extend the domain of force-matched potentials $V(r)$ available only for values
of $r$ restricted by the size of the simulation cell. Fourth, the existence of large
regions of $\bar{\rho},T$ phase space where the Yukawa form of the pair potential
raises the possibility of a semi-analytic theory of a wide range of WDM
systems that are treated by AA, orbital-free-Thomas-Fermi type calculations.

The existence of a simple YFT-type parametrization of pair
 potentials in these WDM systems supports the physical picture of them being
 essentially metallic liquids; this being true even in the case of $l$-C
 or $l$-Si where a different paradigm has often been advanced, within the
 intuitive  chemical picture of covalent bonding. The paradigm of  ``bond-order
 potentials'' and their limitations in WDM applications are now well known.
 Just as the  ``bond-order potentials'' may be used in simulation codes that
 call for repeated evaluations of the potentials, these YFT potentials may also be
 similarly used, much more simply. The method based on constructing  FM potentials
 extracted from more fundamental DFT-MD simulations may
become computationally prohibitive especially at higher temperatures, although the
 underlying physics for $\theta\gtrsim 1$ is Yukawa-like.

 Standard DFT fails when ground states contain significant non-Born-Oppenheimer
 corrections, while excited states may involve conical intersections. 
Nevertheless DFT-MD is  an excellent method for testing 
methods like the NPA, or in providing inputs to reference
data bases for systems where the Born-Oppenheimer approximation
holds and  when experimental data are not available.

\subsection{Pair potentials, pair energies, and the total free energy}
It should be noted that the
 pair potential $V(r)$ and the PDF $g(r)$ are sufficient to evaluate the
 pair-interaction energy  $E_{12}^{int}$ (which is an excess energy referred to the
non-interacting case) or the excess free energy among the ions inclusive of their
 correlation  corrections. But an additional coupling constant integration
 using a potential $\lambda V(r)$ and a PDF $g(r,\lambda)$ with $\lambda$ varying
 from  zero to unity is  needed to obtain the ion-ion contributions $F_{12}$ to
 the free-energy. For instance, the ``excess free-energy" which results from the
 ion-ion pair interaction beyond the mean-field energy is the ion-ion correlation
 energy. This can be written as a coupling-constant integration which may be
 conceptualized as replacing the ionic charge $\bar{Z}|e|$ by $\bar{Z}|e|\lambda$
 where $\lambda$ is the charging parameter~\cite{Onsager33,Kirkwood35} brought up
 to unity from the noninteracting value when $\lambda=0$. The coupling constant
 integration is
  \cite{Morita1960Theory, Yan2004Statistical, PDWXC}
\begin{eqnarray}
\label{coupl.eqn}
F_{12}/\bar{\rho}&=&\int^1_0 d\lambda\bar{\rho}\int 4\pi r^2 dr\frac{V(\lambda,r)}{\lambda}h(\lambda,r),\\
V(\lambda,r)/\lambda &=& V(r),\\ 
h(\lambda,r)&=&g(\lambda,r)-1.
\end{eqnarray}
An actual calculation by this method requires the evaluation of a sequence
 of ion-ion PDFs
$g(\lambda,r)$ with a sequence of potentials $V(\lambda,r)$ for about a dozen
 values of $\lambda$ to carry out the $d\lambda$ integration. Such calculations
 have been done for simple potentials like the electron-electron Coulomb
 interaction in the  context of calculations for the electron exchange-correlation
 energy at finite-$T$~\cite{PDWXC}.
However, if the HNC integral equation is used, then the above
coupling constant integration can be avoided~\cite{Silbert92}. Such a method has been
implemented for warm dense plasmas in Ref.~\cite{eos95} in the context
of the NPA. 

Equation (\ref{coupl.eqn}) shows how the excess free energy of a pair of ions placed
 placed in the medium containing electrons and ions is related to the pair potential.
 The contributions beyond
the mean-field energy contained in the many-body ion-ion interactions is brought in via
their effect on the PDF $g(\lambda,r)$ via the charging process.
 At each charging step, as the interaction grows stronger, the weighting of
 correlations from, say, three-body terms brought in via the Ornstein-Zernike equation
 changes and contributes to the final value of the pair-energy. In the NPA model, the
 basic pair potential $V(\lambda,r)$ defines the direct correlation function $c(r)$
 that enters the Ornstein-Zernike equation. In fact, asymptotically
\begin{equation}
c(\lambda,r)=-V(\lambda,r)/T.
\end{equation}
The NPA's structure, and how to extend it to include core-polarization effects and corrections
 to the kinetic energy functional that contribute to the pair potential etc., are
 described in Sec.~2 of Appendix B of Ref.~\cite{eos95}. It does not include chemical
 bonding. In fact,  persistent molecular ions arising from chemical bonding are
compact entities that should be regarded as additional species present in the mixture.

The $F_{12}$ determines the ion-ion contributions to the internal energy, pressure,
 specific heats and other thermodynamic quantities. The coupling-constant integration
 can be explicitly carried out, say, for a dozen values of $\lambda$. Usually, even
with such coupling constant integration, the NPA calculation of the free energy is
 computationally not heavy. However, we have found this coupling-constant integration, 
i.e., the $\lambda$-integration unnecessary as it can be 
 side-stepped within the HNC integral-equation approaches,
 as discussed in Ref.~\cite{eos95},  or by using reference hard-sphere-fluid
 models~\cite{Pe-Be,perrotRPA91}.
 
In effect, we may write the total Helmholtz free energy $F$ in terms of
 the contributions arising from the
electrons, ions, and their interactions in the form~\cite{Pe-Be,eos95}:
\begin{equation}
\label{helm.eqn}
F=F^0_e+F^{xc}_e+F_{em}+F_{12}+F^0_I.
\end{equation}
The first two terms deal with the free energy of the non-interacting
uniform electron fluid and its finite-$T$ exchange-correlation energy
 at the given density and temperature $T$. The last term, $F^0_I$ is the
ideal (classical) free energy of the ion subsystem. The zero of
energy may be assumed to be the infinitely separated set of nuclei and electrons,
when the energy $E_A$ of $\bar{\rho}$ isolated atoms per unit volume (obtained from
a standard atomic physics calculation) must be added to Eq.~(\ref{helm.eqn}).
This term eliminates itself in calculations of quantities like the pressure and specific heat
 as only differences in the free energy (derivatives) arise.  

The third term, $F_{em}$ of Eq~\eqref{helm.eqn} is the embedding energy of the neutral pseudo-atom
 NPA in the uniform electron fluid.
The fourth term, $F_{12}$ contains the interactions between unscreened
pseudoatoms brought in via the pair potential, PDF, and the
ion-ion correlation effects as already discussed in the context of
the coupling-constant integration.

Given the average number of free electrons
per ion (viz., $\bar{Z}$), the first three terms $F^0_e, F^{xc}_e$, and $F^0_I$ at the
 given temperature are trivially evaluated. If an NPA calculation
 or an equivalent AA calculation is available, $F_{em}$ and 
$F_{12}$ are also available. Otherwise, approximate ``universal'' embedding energy functionals
at finite-$T$ are available for use~\cite{Zaremba85}. Hence, if parametrized pair potentials
are available, classical MD methods may be used to calculate $F_{12}$, completing
the evaluation of the full Helmholtz energy $F$, with respect to the reference zero of energy
which may be conveniently taken to be the limit of infinitely separated neutral atoms,
or alternatively, infinitely separated nuclei and electrons. Such calculations of the
total free energy, internal energy and pressure, and electrical conductivity
 for Al, Be, C, H, Li, Si, etc., have been carried out in previous publications 
using the NPA method.
\subsection{Non-uniform systems}
Consider a plasma with a density or temperature gradient
along the $x$-direction; if the gradient region can be replaced by a sequence
of uniform slabs, with a thickness not less than the correlation-sphere radius
 $R_c(x)$ in the direction of the
density gradient, and a ``thermal-energy thickness" of not more than
 $\bar{Z}(x)/r_{\rm ws}(x)$
 in the direction of the temperature gradient, then the density and temperature dependent
 pair potentials of each slab
can be used in the term $F_{12}$ of Eq.~(\ref{helm.eqn}). The other terms in Eq.~(\ref{helm.eqn}) are less
complex and can be easily evaluated or interpolated using the density profile. Of course, this
assumes that the varying density and temperature region does not traverse a phase transition or
cause significant changes in the nature of the ions and hence the pair potential does not
change strongly. Harbour {\textit{et al.}}.~\cite{HarbEOSPhn17}
 found surprisingly good results even when the method
was applied to the near-surface region of an Al slab.  

In the purely classical MCP approach, the expectation is
 that such gradients in temperature, density
and ionization state are already covered by the ML
 algorithms that generated the potentials; hence a partitioning of the
 gradient region and quantum calculations for each of them
are avoided. In many cases, when dealing with density
or temperature gradients, doing a full NPA calculation for each density
 slab is quite easily done.

\subsection{Mixtures of ions}
\label{mixture.sec}
We briefly bring out the implications of the availability of parametrized
potentials in dealing with mixtures. A more complete discussion of the NPA
approach to mixtures is found in Ref.~\cite{eos95}.

If we consider a simple mixture of $n_s$ species of ions, we need to specify 
$(n_s+1)n_s/2$ different  pair potentials that can all be specified
 using only $n_s$ pseudopotentials (see Eq.~\eqref{uei.eqn}) and the average
electron density (or, equivalently, $\bar{Z}$ of the mixture) which specifies
 the electron response function.  However,  a number of concerns have to
 be resolved before the theory can be exploited.

For example, if the WDM mixture is between two different ions having the same $\bar{Z}$, e.g., WDM Si
 and WDM C for which $\bar{Z}$ = 4 holds for a wide range of densities and temperatures,
 then the calculation of pair potentials and thermodynamics of such mixtures becomes trivial
 as the ionization balance is preserved.

 However, if we consider a a sample  of normal density ($\bar{\rho}$ = 2.7 g/cm$^3$) WDM
Al at any temperature up to about $T$ = 12 eV ($\sim E_F$),
  $\bar{Z}_{\rm Al} \approx 3$. We may
isochorically replace a fraction of the atoms with Li atoms, to give composition
 fractions $x_{\rm Al},x_{\rm Li}=1-x_{\rm Al}$, with a mixture density 
$\bar{\rho}_m=x_{\rm Al}\bar{\rho}_{\rm Al}+x_{\rm Li}\bar{\rho}_{\rm Li}$,
Unfortunately, it does not necessarily follow that 
$\bar{Z}_m=x_{\rm Al}\bar{Z}_{\rm Al}+x_{\rm Li}\bar{Z}_{\rm Li}$. If this is so, when 
$x_{\rm Al} = x_{\rm Li}$ = 0.5, then  $\bar{Z}_m=$ 2.
One may perhaps envisage a preliminary AA calculation where
all ions will be replaced by a doubly ionized O atom but that model
 may well be too approximate. If $\bar{Z}_m=$ 2 holds, the relevant
 pseudopotentials and pair potentials will be those appropriate to
  Al$^{2+}$ and Li$^{2+}$
 ions that we can obtain from the appropriate NPA calculations rather
 than for an O-like average atom. 
In reality, the 1$s$ electrons of the Li$^{1+}$ have such a high level
of stability that such ionization to double-ionized Li will be energetically
 strongly unfavorable unless  the temperature of the system is
high enough. At lower temperatures, the mixture will become a multi-species plasma with singly
 ionized Li and several types of Al ions with ionization states of one, two, and three
 to provide a mean ionization of $\bar{Z}_m$ equal to two. The vector of composition
 fractions $\{x_s\}$ is such
 that it minimizes the free energy given by Eq.~(\ref{helm.eqn}), and $x_s$ satisfies
 $(\partial F/\partial x_s)=0$. Such conditions can be imposed as Lagrange multipliers in
codes written for more automated calculations on mixtures.
Such first-principles determinations of $x_s$ as well as
 the pair potentials  and thermodynamics of a system inclusive of multiple ionization states
 have been demonstrated  using the NPA in Ref.~\cite{eos95}. A knowledge of the vectors $\{x_s\}$ and
$\{\bar{Z}_s\}$ is useful for understanding the spectra and line broadening in such mixtures.
In fact plasma spectroscopy can validate the values of $\{\bar{Z}_s\}$ and $x_s$ predicted via such NPA
calculations.

A QMD calculation for a mixture of ions requires a non-trivial effort.
It will yield the thermodynamic data but not the composition fractions
$\{x_s\}$ nor the charge states of the different species of ions $\{\bar{Z}_s\}$. The pair potentials
 including the cross-species potentials can in principle be extracted using force matching to such
 QMD calculations. If only very small admixtures of an impurity atom are considered, the major component
 will retain its $\bar{Z}$ and the calculations via the NPA greatly simplify. However, such extreme
 concentration fractions is computationally challenging for QMD simulations as there has to be sufficient
minority-species atoms in the simulation for statistical accuracy.

\section{Conclusion and Outlook}
We have presented an analytic form for ion-ion pair potentials for cold-, warm- to hot-dense-matter
systems having a significant degree of ionized electrons, as found in metals and plasmas.
 Our model uses a Yukawa-like pair potential augmented by
 a thermally damped oscillatory Friedel tail, where the parameters in the potential are obtained from
a single-center Kohn-Sham calculation within the NPA model or from QMD calculations. The YFT model is applicable
 at arbitrary coupling, unlike the standard Yukawa form valid only in the high-temperature high-density
limit such that $\theta\gtrsim 1$. We  have shown that it accurately reproduces results from QMD simulations
for a wide range of plasma conditions from near the melting point to high temperatures at densities
typically found in liquid metals and plasmas of interest in many areas of study.

Using reliable pair interaction potentials such as those from force matching
 to $N$-atom DFT-MD data, we find that in all cases studied here, the YFT agrees closely with them
and reproduces structural quantities from DFT-MD with good accuracy. The YFT model is based on a
 linear pseudopotential that is consistent
 with the NPA calculation as well as its sophisticated ionization-balance calculations that go into
the evaluation of the mean ionization per atom. The latter is given by
$\bar{Z}$ and is self-consistently constrained to satisfy the Friedel sum rule. 
The pseudopotential and the mean ionization $\bar{Z}$  are primary inputs to transport
 calculations that are not easily available from $N$-atom DFT-MD or from orbital-free DFT calculations.
 The computational costs of NPA calculations, or YFT fits are
 orders of magnitude less than for QMD where millions of $N$-center DFT calculations for
 millions of ionic configurations are needed. We also note that in contrast to $N$-atom orbital-free
 molecular dynamics, NPA calculations (i) are computationally faster, (ii) do not suffer from
 finite-size effects, (iii) decrease statistical errors associated with short-time simulations, 
(iv) statistically resolve all species in systems with impurities, 
(v) are computable at any temperature, and (vi) generate pseudopotentials from their own
all-electron calculations instead of requiring them as external inputs.

In the context of FM potentials, the YFT model provides a way to correct for finite-size effects;
since it  can be evaluated at small and large interparticle distances where as extrapolation is
 needed for FM potentials obtained from finite-cell
calculations.

While the NPA model in its current formulation assumes the
 existence of a weak pseudopotential for which linear response theory is applicable,
 the YFT model is able to also accurately represent potentials obtained
by force matching to systems like low-density $l$-C where NPA methods prove inadequate.
 The NPA fails  if the system has a low density of free electrons (e.g., expanded metals
 and attenuated plasmas) or when no free-electrons what so ever (e.g., as in liquid molecular
hydrogen) are present.  In such situation, physics-informed ML methods are useful as they generate
 a very detailed description of the physical system
 including ion-ion many-body interactions and bonding descriptions inclusive of angular
correlations. However, results from such ML methods as well as $N$-atom DFT
 methods should be validated against experimental data or strictly {\it ab initio} methods
 that directly obtain the wave function, as in
 configuration-interaction calculations, coupled-cluster calculations or quantum
 Monte Carlo and path-integral methods, when they are feasible. 

 While this work is focused on single-component systems, a natural extension of this work is to
 study plasma mixtures of different elements, or mixtures of different charge states
of the same element~\cite{eos95}. The possibility of treating mixtures would
 allow for the calculation of the equations of state, charge states of ions for  plasma spectroscopy,
 and plasma transport  properties like interdiffusion coefficients,
that are of interest in hydrodynamic models used in simulations of high-energy-density
 experiments.

\section*{ACKNOWLEDGMENTS}
The authors would like to thank Raymond C. Clay III for useful discussions.
L.J.S. and M.S.M. acknowledge support from the U.S. Air Force Office of
Scientific Research Grant No. FA9550-17-1-0394.


\newpage
$\;$
\appendix
\section{\label{sec:bound_free}The division of the total electron density}
In this paper we have studied C, Al and Li in their fluid-like  WDM states 
from temperatures slightly above their melting points, and for a range of  high temperatures
extending to  $T = $ 100 eV, using the NPA model of Perrot
and Dharma-wardana. We have presented the theory of the NPA model based on the division
of the total electron density $n(r)$ around each nucleus into a bound-electron
distribution $n_b(r)$ and a free electron distribution $n_f(r)$. 

This division is not peculiar to the NPA, but is in common use
 in computational physics of materials. It is implicit in the
 recognition of the existence of band electrons and core
electrons in the theory of solids, and in the recognition of
 the existence of valence electrons and core electrons in the
 theory of chemical bonding. Furthermore, casting this division
 into a formal scheme is already found in the theory of
 pseudopotentials of the elements that are used in QMD codes
 as a means of avoiding all-electron atomic calculations. Hence
 we have more than half a century of experience (from the theory
 of pseudopotentials) in dealing with this division of the total
 electron density into core electrons that are bound to the
nucleus, and ``free electrons". However, we review the matter further
in the context of WDM studies, and how this division is implemented
in the  NPA model.

This division is unambiguous when
the bound electron density obtained from the finite-$T$ Kohn-Sham calculation
is well contained inside the Wigner-Seitz sphere of the
ion. That is:
\begin{equation}
\label{bf-div.eqn}
\begin{cases}
    n(r) = n_b(r)+n_f(r), \quad  &r<r_{\rm ws}\\
    n_b(r) = 0, \;  &r>r_{\rm ws}.
\end{cases}
\end{equation}
The spin index has been summed over in the above equation but  applies to spin-polarized
DFT-type calculations as well as to unpolarized calculations. An immediate consequence of the
 above equation is a simple result for
 the net charge $\bar{Z}$ on an ion of nuclear charge $Z_n$ when immersed in a plasma.
 If the nucleus acquires $n_b$ bound electrons, given by the space integral of $n_b(r)$, then,
when Eq.~\eqref{bf-div.eqn} holds we can write
\begin{equation}
\bar{Z}=Z_n-n_b. 
\end{equation}
The claim that the use of such a $\bar{Z}$ is not possible because there is ``no operator
in quantum mechanics'' that corresponds to an observable such as the ``mean ionization'' is not valid
in much the same sense that there is no quantum operator that corresponds to the temperature, or to
the chemical potential of a system. The temperature, and the chemical potentials are introduced
into the theory of quantum statistical systems as Lagrange multipliers to conserve 
energy and particle numbers respectively, within the grand canonical ensemble. The mean ionization
is a Lagrange multiplier associated with the charge neutrality of the system~\cite{DWP82}. Furthermore,
 an operator for $\bar{Z}$ can be constructed  via a scattering operator whose structure corresponds
 to the formulation of the Friedel sum rule. 

Experimentally $\bar{Z}$ is routinely measured
in laboratory plasmas directly using Langmuir probes, or indirectly via various optical
 methods like
X-ray Thomson spectroscopy. Furthermore, the  NPA calculation is
a thermodynamic calculation for the free energy, and provides the sophisticated many-body
 generalization
of the Saha equation from the ionization balance where many-body effects are not treated or treated
only in the weak-coupling limit.

Aside from such misplaced formal objections about  ``a lack of a quantum operator for $\bar{Z}$,"
there are other more practical difficulties with Eq.~(\ref{bf-div.eqn}).
In many cases, the electron density associated with a bound state close to its
pressure-ionization threshold, or to its temperature ionization window is not fully contained
 inside the Wigner-Seitz sphere. The bound density may still have a substantial magnitude
  outside the Wigner-Seitz sphere even if the density were exponentially decaying, or the
 density might appear in the form of a continuum resonance.
Then the classification as a ``bound density,'' or a ``free density'' requires additional
 clarifications since Eq.~\eqref{bf-div.eqn} becomes inadequate as it stands. 

Any bound electrons that spread outside the Wigner-Seitz sphere need to be shared by other ions in the
 system and a method of redistribution (e.g.,
based on multiple-scattering among ``muffin tins," as used in the Korringa-Kohn-Rostoker model),
or based on other 
considerations becomes necessary. However, if the coordination number is, say eight, while the
charge outside the Wigner-Seitz cell is shared by eight neighbours, they in turn contribute a similar
charge into the Wigner-Seitz cell, and they need to be explicitly included in the Kohn-Sham calculation
if the effects are important, as may be the case in very high-density systems.

Similarly, a ``free'' electron, i.e., an electron in a continuum state that spends a considerable
 amount of time localized near a nucleus is a
 ``virtual bound state'', as first identified in studies of the electronic states of
 impurities in metals  by Friedel~\cite{Friedel1962}. The Friedel sum rule unambiguously defines
 the division of the charge density into a bound part and a free part. This is carried out via
a phase-shift analysis of the continuum solutions of the Kohn-Sham equations for the electron eigenstates
of the ion immersed in its host medium. The Friedel
sum rule, given originally for the $T=0$ case by Friedel was generalized by Dharma-wardana and
 Perrot to finite temperatures~\cite{DWP82} and used in the NPA in the self-consistent determination
 of $\bar{Z}$ so as to satisfy the sum rule. 

In cases where a bound state extends significantly to regions $r>r_{\rm ws}$,
further steps need to be taken to ensure that the sum rules (e.g., Friedel sum rule, $f$-sum rule) etc., 
are accurately satisfied  (i.e., to $\sim 99\%$) in iteratively solving the
Kohn-Sham equations of the NPA model. In such cases, and in the case of many transition-metal ions,
the bound state that extends out of the Wigner-Seitz sphere usually corresponds to cases where a non-integer
$\bar{Z}$ arises in the  NPA calculation. Here a single AA is being made
 to represent a mixture of ions of integer charge. Our experience is that such cases behave as more compact
 ionic systems, i.e., with the bound states well within the Wigner-Seitz sphere,  when the system is
 treated as a mixture of integer-ionized NPAs.

In Fig.1(b) of Ref.~\cite{cdw-Carbon10E6-21}, the extension of the only bound
state of highly compressed C at $T$ = 100 eV (i.e. the $1s$ state)  is considered using the NPA model. 
It is found that for densities beyond $\bar{\rho}$ = 341 g/cm$^3$, i.e., some 100 times the diamond density,
the bound density is not fully contained within the Wigner-Seitz cell. In such cases, the
satisfaction of the sum rules as well as the convergence of the NPA equations becomes poor unless
suitable steps are taken. These steps include (i) inclusion of explicit self-interaction corrections, 
(ii) treatment of the system as a mixture of integer-ionized species instead of an ionization state
which has a fractional part, and (iii) possible use of a subdivision of the charge that extends outside
the Wigner-Seitz sphere in terms of the local coordination number as implied  by the PDF
of the fluid. For instance, Perrot has (as a practical measure) used a somewhat intuitive
 ``cutting function'' approach based on such ideas, without actually deriving it from density-functional
 arguments, in his NPA calculations of WDM Be~\cite{Pe-Be}. A more  rigorous DFT approach would be to
 include such effects in the ion-ion XC functional as well as to use the explicit $g(r)$ to simulate
the field-ion density instead of using a Wigner-Seitz spherical cavity. Another practical approach
 is to solve for the bound part of the NPA Kohn-Sham equations with
the bound-states to have vanishing density at the Wigner-Seitz sphere surface, while the free
 electron density reaches the mean free-electron density only at the edge of the correlation
 sphere where $r=R_c$.

The regions of temperature and density where such special steps need to be taken occur
 for essentially all nuclei carrying bound-electron sates, especially when a weak bound state is
 formed, or when such a weakly bound state disappears into the continuum. However, these special
 regions are far less common than regions where the simpler approach is applicable.
 Our experience is that the EOS data for quantities like the pressure and energy can be smoothly interpolated
 across such ``difficult regions," unless it turns out that a phase transition or some such
 special phenomenon is also occurring in such regions. This is usually easily
detected by monitoring the evolution of the compressibility ratio of the system,
 i.e., the ratio of the interacting compressibility to the ideal gas compressibility. This can be
 evaluated from EOS data, or from the $k\to 0$ limit of the ion-ion structure factor $S(k)$.
 If rapidly changing values of $S(k\to 0)$ and discontinuities are found, 
a more careful investigation is needed, as in Ref.~\cite{cdw-carb22}. 
As such, we have included the compressibility
 ratio as well as the  electrical conductivity in our tabulations given in Appendix~\ref{sec:tabulations}.

\section{\label{sec:tabulations} YFT parameters for  pair potentials of C, Al, Li up to 100 eV}

This section  provides tabulations of YFT fit parameters for the NPA pair-potentials
for fluid WDM states of C, Al, and Li. The Yukawa limit of the YFT 
fit to the NPA potentials are explicitly given as separate tabulations up to $T=$ 100 eV at the normal density.
\subsection{YFT parameters for carbon pair potentials}

Table~\ref{Ypara-C.tab} complements the YFT parameter for C pair potentials
 given in Table~\ref{para-C.tab}, and considers the Yukawa limit of the YFT parametrization.

\begin{table*}
\caption{\label{Ypara-C.tab} The parameters of the YFT potential fit to the NPA potentials for
C at the graphite density of $\bar\rho =$ 2.270 g/cm$^3$. The fit reduces to a
simple Yukawa potential at sufficiently high temperatures. The compressibility ratio given by $k \to 0$
limit of the structure factor, $S(0)$ obtained from the full NPA potential, and the static electrical
conductivity $\sigma$ are also given in each case
as a reference thermodynamic datum and a transport datum.
}
\begin{ruledtabular}
\begin{tabular}{lcccccccc}
$T$ [eV]            &  20      &  30     &  40      &  70     &  100 \\
\hline
$\theta$      & 0.9268   & 1.390   & 1.852    & 3.241   & 4.634   \\
$\bar{Z}$           & 4.0000   & 4.0004  & 4.0060   & 4.0050  & 4.1311  \\
$S(0) $             & 0.3221   & 0.28591 & 0.2617   & 0.2125  & 0.18436   \\
$\sigma$[10$^6$S/m] & 0.5714   & 0.6666  & 0.8006   & 1.433   & 4.049   \\
\hline
$a_y$               & 21.1745  & 13.381   & 10.0166  & 6.38984  & 5.48699   \\
$k_y$               & 1.09593  & 0.87968  & 0.753483 & 0.560216 & 0.480174   \\
\end{tabular}
\end{ruledtabular}
\end{table*}

\newpage
$\;$

\subsection{YFT parameters for aluminum pair potentials}

The YFT parameters for Al potentials from the NPA model for the low $\theta$ region are given
in Table~\ref{para-Al.tab} for several densities, while Table~\ref{Ypara-Al.tab} is for
Al at the normal density from $T = $ 8 eV to 100 eV where the Friedel tail may be neglected.

\begin{table*}
\caption{\label{para-Al.tab}The parameters of the YFT
potential for $V(r)/T$ for Al with $r$ in atomic units.
  The fits are for approximately the range of
 $1<r/r_{\rm ws}<4$ and covers up to $\sim 10$ units in the energy scale of $V(r)/T$.
 The NPA potentials calculated via Eq.~(\ref{pair.eqn}) are used for the fit.
}
\begin{ruledtabular}
\begin{tabular}{lcccccccc}
$\bar{\rho}$ [g/cm$^3]$   
                & 2.0   &  2.0       &  2.7      &  2.7     &  4     &  4    & 5          &  5 \\
\hline
$T$[eV]         &   1     &   2       & 1        & 2        &  1     &  2    & 1          &   2  \\
\hline 
$a_{\rm y}$     & 6509.14 & 3519.59   & 6655.83  & 2370.2 & 4643.04  & 2175.91 & 4384.8   & 2084.72 \\
$k_{\rm y}$     & 1.82023 & 1.83385   & 1.82532  & 1.71056 & 1.73393 & 1.70921 & 1.73697  & 1.71774 \\
$a_{\rm ft}$    & -20.6658 & -9.22571 & 14.8642  & 6.09412 & 11.2754 & 4.78417 & 10.6638  & 4.57885 \\
$k_{\rm ft}$    & 0.0      & 0.0      & 0.0      & 0.0     & 0.0     & 0.0     & 0.0      & 0.0    \\
$q_{\rm ft}$    & 1.73594 & 1.76649   & 1.98027  & 1.99726 & 2.13319 & 2.17356 & 2.19991  & 2.21888 \\
$\phi_{\rm ft}$          & -8.87522 & -9.15885 & -6.72432 & -6.69006& -6.87101& -7.0183 & -6.84845 & -6.88636 \\
\end{tabular}
\end{ruledtabular}
\end{table*}

\begin{table*}[h]
\caption{\label{Ypara-Al.tab} The parameters of the YFT potential that reduces
to the Yukawa model applicable for normal
density ($\bar{\rho}=2.7$ g/cm$^3$) Al, $r_{\rm ws}=$2.98997 a.u.,
for $\theta>0.68 $ for $V(r)/T=a_y\exp(-k_yr)$
  with $r$ in atomic units. The fits are for approximately the range of $1<r/r_{\rm ws}<4$.
 The compressibility ratio given by $k \to 0$ limit of the
structure factor, $S(0)$ obtained from the full NPA potential
as a reference thermodynamic datum. A full discussion of the static electrical
conductivity of Al is available in Ref.~\cite{cond3-17} for this range of temperatures
and density.
}
\begin{ruledtabular}
\begin{tabular}{lccccccc}
$T$ [eV]            & 8       &  10       &  20      &  30     &  40     &  70        &  100 \\
\hline
$\theta$      & 0.6857  &  0.8545   &  1.549   &  2.024   & 2.413    & 3.481    & 4.567   \\
$\bar{Z}$           & 3.0029  &  3.0164   &  3.495   &  4.299   & 5.085    & 6.796    & 7.7205   \\
$S(0) $             & 0.23786 &  0.24484  &  0.23981 &  0.21329 & 0.18825  & 0.15564  & 0.082849 \\
$a_y$               & 136.648 &  77.8178  &  25.6476 &  21.6325 & 20.4458  & 19.1447  & 17.6433  \\
$k_y$               & 1.19013 &  1.0532   & 0.756828 & 0.679670 & 0.632411 & 0.54239  & 0.461407 \\
\hline
\end{tabular}
\end{ruledtabular}
\end{table*}

When $\theta$ exceeds $\sim 0.7$ the pair potentials
of Al at the normal density of $\bar\rho = $ 2.7 g/cm$^3$ can be fit to a simple Yukawa
form with the Friedel tail neglected. If the YFT fit is retained even at these temperature,
 the fit improves for large-$r$ but has little impact on the PDFs,
 except possibly for the $k\to 0$ limit of $S(k)$. Hence the $S(0)$
 calculated with the full  NPA potential is also given in Table~\ref{Ypara-Al.tab}.


\subsection{YFT parameters for lithium pair potentials}

The YFT parameters for Li-Li potentials derived from NPA are given in Table~\ref{para-Li.tab}, while
the Yukawa form which is sufficient at higher temperatures is given in Table~\ref{Ypara-Li.tab}.

\begin{table*}[h]
\caption{\label{para-Li.tab}The parameters of the YFT
potential for $V(r)/T$ for Li with $r$ in atomic units.
The fits are for approximately the range.
 $1<r/r_{\rm ws}<4$ and covers up to $\sim 10$ units in the energy scale of $V(r)/T$.
 The NPA potentials calculated via Eq.~(\ref{pair.eqn}) are used for the fit. 
}
\begin{ruledtabular}
\begin{tabular}{lccccccccc}
$\bar{\rho}$ [g/cm$^3]$
              & 0.513   &  0.513   & 0.513   &  0.6     &  0.6    &  0.85   &  0.85   & 1.0   & 1.0 \\
\hline
$T$[eV]       &  0.054  & 1.0      & 4.5     & 0.054    &  1.0    &  0.054  &  1.0    & 0.054    & 1.0  \\
\hline 
$a_{\rm y}$   & 11603.5 & 48.0952  & 11.0273 & 20382.5  & 91.2908 & 44733.8 & 103.144 & 10809.8 & 131.555  \\
$k_{\rm y}$   & 1.56916 & 0.92618  & 0.729692& 1.72629  & 1.16503 & 2.19708 & 1.31930 & 1.59543 & 1.2544 \\
$a_{\rm ft}$  & 298.993 & 19.1008  & 0.0     & 242.327  & 17.4002 & 418.188 & 31.2169 & 151.662 & 9.53177 \\
$k_{\rm ft}$  & 0.0     & 0.0      &    -    & 0.0      & 0.0     & 0.10317 & 0.171479& 0.0     & 0.0 \\
$q_{\rm ft}$  &-0.918337& 1.01096  &    -    & 0.996746 & 1.00526 & 1.15036 & 1.12187 & 1.20109 & 1.17779 \\
$\phi_{\rm ft}$     & 8.84561 & 9.69025   &    -  & 9.59544  & 9.667156& 8.91247 & 9.08969 & 9.05297 & 9.22377 \\
\end{tabular}
\end{ruledtabular}
\end{table*}

\begin{table*}[h]
\caption{\label{Ypara-Li.tab} The parameters of the YFT potential fit to the NPA potentials for
Li at the normal density of $\bar\rho$ = 0.513 g/cm$^3$. The fit reduces to a
simple Yukawa potential at sufficiently high temperatures. The compressibility
 ratio given by $k \to 0$ limit of the
structure factor, $S(0)$ obtained from the full NPA potential, and the static electrical conductivity
$\sigma$ are also given in each case as a reference thermodynamic datum and a transport datum.
}
\begin{ruledtabular}
\begin{tabular}{lcccccccc}
$T$ [eV]            & 10     &  20      &  30      &  40     &  70      &  100 \\
\hline
$\theta$      & 2.0743 & 3.188    & 4.045    &  4.902  & 7.706    & 10.72   \\
$\bar{Z}$           & 1.0800 & 1.604    & 2.061    &  2.378  & 2.793   &  2.905  \\
$S(0) $             & 0.4989 & 0.4059   & 0.3452   &  0.3096 & 0.2668   & 0.2539   \\
$\sigma$[10$^6$S/m] & 1.0400 & 1.0962   & 1.1886   &  1.3657 & 1.9959  & 1.3679 \\
\hline
$a_y$        & 4.27453  & 3.76352  & 3.93123  & 3.86079   & 3.04285  & 2.32109   \\
$k_y$        & 0.540474 & 0.434532 & 0.393156 & 0.361723  & 0.293366 & 0.249578 \\
\end{tabular}
\end{ruledtabular}
\end{table*}

The Yukawa limit of the YFT fits is given in the table~\ref{Ypara-Li.tab}, for Li at the normal density
of $\bar\rho$ = 0.513 g/cm$^3$.

$\;
\;
\;
$

\newpage
$\;$

\end{document}